\documentclass[aps,groupedaddress,nofootinbib]{revtex4-2}

\usepackage[english]{babel}
\usepackage[utf8]{inputenc}
\usepackage{amsmath}
\usepackage{mathbbol}
\usepackage{amssymb}
\usepackage{bbold}
\usepackage{graphicx,amsfonts}
\usepackage[colorlinks=true,
linkcolor=red,
urlcolor=blue,
citecolor=red,linktocpage]{hyperref}
\usepackage{bm}
\usepackage{mathrsfs}
\usepackage{enumerate}
\usepackage{amsthm}
\usepackage{bbm}
\usepackage{comment}
\usepackage{physics} 

\begin{document}
	
\title{Black-hole spectroscopy: quasinormal modes, ringdown stability and the pseudospectrum}
	
\author{Kyriakos Destounis$^{1,\,2}$ and Francisco Duque$^{3,\,4}$}
\affiliation{$^1$Dipartimento di Fisica, Sapienza Università di Roma, Piazzale Aldo Moro 5, 00185, Roma, Italy}
\affiliation{$^2$INFN, Sezione di Roma, Piazzale Aldo Moro 2, 00185, Roma, Italy}
\affiliation{${^3}$CENTRA, Departamento de F\'{\i}sica, Instituto Superior T\'ecnico -- IST, Universidade de Lisboa -- UL, Avenida Rovisco Pais 1, 1049 Lisboa, Portugal}
\affiliation{${^4}$Max Planck Institute for Gravitational Physics (Albert Einstein Institute) Am Mühlenberg 1, D-14476 Potsdam, Germany}

\begin{abstract}
Black-hole spectroscopy is a powerful tool to probe the Kerr nature of astrophysical compact objects and their environment. The observation of multiple ringdown modes in gravitational waveforms could soon lead to high-precision gravitational-wave spectroscopy, thus it is critical to understand if the quasinormal mode spectrum itself is affected by astrophysical environments, quantum corrections, and other generic modifications. In this chapter, we will review the black-hole spectroscopy program and its challenges regarding quasinormal mode detection, the overtone status and the recent evidence that supports the existence of nonlinearities in the spectrum of black holes. We will then discuss a newly introduced non-modal tool in black-hole physics, namely the pseudospectrum; a mathematical notion that can shed light on the spectral stability of quasinormal modes, and discuss its novel applications in black holes and exotic horizonless compact objects. We will show that quasinormal modes generically suffer from spectral instabilities, explore how such phenomena can further affect black-hole spectroscopy, and discuss potential ringdown imprints and waveform stability issues in current and future gravitational-wave detectors.
%
\end{abstract}
\maketitle
\tableofcontents
\clearpage
%

\section{Introduction}

The observation of gravitational waves (GWs) from compact binary sources has become a primary avenue of scientific exploration. GW detections have become routine, leading to a novel era of GW astrophysics~\cite{Barack:2018yly}. Systematic multimessenger observations~\cite{Meszaros:2019xej} from the Laser Interferometer Gravitational-Wave Observatory (LIGO), Virgo and KAGRA~\cite{LIGOScientific:2020ibl,LIGOScientific:2021djp}, as well as future ground- and space-based interferometers, such as the Laser Interferometer Space Antenna (LISA)~\cite{LISA:2017pwj,Amaro-Seoane:2022rxf}, Taiji~\cite{Hu:2017mde} and TianQin~\cite{TianQin:2015yph,Gong:2021gvw}, will unequivocally strengthen our understanding of black holes (BHs) and strong-field gravity~\cite{Amaro-Seoane:2022rxf,LISA:2022kgy}. GWs carry fundamental information regarding the remnant's externally observable quantities, their surrounding environment~\cite{Leung:1997was,Barausse:2014tra,Bamber:2021knr,Cardoso:2021wlq,Cardoso:2022whc,Destounis:2022obl} and their potential deviations from general relativity (GR)~\cite{Clifton:2011jh,Yunes:2013dva}. 

According to uniqueness theorems in GR~\cite{PhysRevLett.26.331}, the merger of two isolated BHs eventually leads to a stationary compact object described by at most three parameters: mass, charge (which is astrophysically expected to be negligible) and angular momentum. Due to the dissipative nature of GWs, binaries typically undergo three distinct stages, before forming a final, stable configuration, namely the inspiral, merger and ringdown. We are now able to accurately simulate binaries and extract fundamental information regarding the initial parameters of their constituents, such as their respective masses and spin, as well as the externally-observable properties of the remnant. Such simulations are usually quite cumbersome and demand the use of powerful numerical tools~\cite{Lehner:2001wq}. Nevertheless, for the inspiral and ringdown stages we can take advantage of semi-analytic models which agree surprisingly well with numerical relativity simulations. During the inspiral stage of extreme-mass-ratio systems~\cite{Barack:2018yvs, Wardell:2021fyy, Albertini:2022rfe,Destounis:2020kss,Destounis:2021mqv,Destounis:2021rko,Destounis:2023gpw}, as well as during the ringdown stage, BH perturbation theory provides the most appropriate schemes to treat the evolution of these scenarios. 

In particular, by simply analyzing the ringdown signal of perturbed BHs and compact objects, through an infinite (discrete) set of vibrational spectra called quasinormal modes (QNMs)~\cite{Kokkotas:1999bd,Berti:2009kk,Konoplya:2011qq}, one can extract paramount information regarding the nature of the final remnants; a task called BH spectroscopy~\cite{Berti:2005ys}. So far, there are various claims that besides the fundamental QNM, several overtones have been detected from gravitational waveforms, that are critical for parameter estimation and validity tests of GR~\cite{Giesler:2019uxc,Isi:2019aib,Isi:2020tac,Isi:2021iql,Franchini:2023eda}. Nevertheless, we are not yet entirely certain if the BH spectroscopy program is suitable for non-vacuum objects residing in astrophysical environments~\cite{Leung:1997was,Barausse:2014tra,Bamber:2021knr,Cardoso:2021wlq,Cardoso:2022whc,Destounis:2022obl}, quantum BHs with horizon-scale corrections~\cite{Abedi:2020ujo}, exotic compact objects~\cite{Cardoso:2016rao,Cardoso:2016oxy,CardosoPani2019} or novel configurations that incorporate modifications of gravity~\cite{Blazquez-Salcedo:2017txk,Destounis:2018utr,Moulin:2019ekf,Blazquez-Salcedo:2020caw,Pierini:2021jxd,Vlachos:2021weq,Chatzifotis:2021pak,Ghosh:2023kge}; ingredients that are widely present in the Universe and can affect the QNM spectrum significantly~\cite{Nollert:1996rf,Nollert:1998ys,Jaramillo2020,Jaramillo:2021tmt,Jaramillo:2022kuv,Destounis:2021lum,Cheung:2021bol,Boyanov:2022ark,Berti:2022xfj,Ghosh:2023etd}. Recent disputes regarding the detection of overtones~\cite{Cotesta:2022pci,Isi:2022mhy}, as well as the presence of nonlinear QNMs~\cite{Cheung:2022rbm,Mitman:2022qdl,Baibhav:2023clw,Yuste:2023,Nee:2023osy} in numerical simulations can pose significant limitations in GW data analysis and BH spectroscopy. To exacerbate on the aforesaid issues, very recent developments on BH QNM instabilities~\cite{Jaramillo2020,Jaramillo:2021tmt,Destounis:2021lum,Jaramillo:2022kuv,Boyanov:2022ark,Konoplya:2022pbc}, when various external perturbations to the vacuum solutions are taken into account, may also affect overtones and even the fundamental mode~\cite{Cheung:2021bol,Berti:2022xfj,Courty:2023}.

More specifically, consider that a perturbation is added to the BH effective potential. If the new perturbed QNMs of the dirty BH have disproportionately migrated to the complex plane with respect to the scale of the perturbation introduced to the effective BH potential, then the BH QNMs exhibit a \emph{spectral instability}. Spectral instabilities are crucial in what regards the QNM overtone status and can be quantified by employing the mathematical notion of the pseudospectrum~\cite{Trefethen:2005}. It is the aim of this chapter to discuss contemporary non-modal tools of perturbation theory, describe how to employ those novel schemes in order to shed more light on the nature of ringdown physics, and to further discuss the underlying spectral features of perturbed compact objects that may limit the goals of BH spectroscopy and ringdown waveform modeling.

\section{Black-hole perturbation theory and quasinormal modes}

When binaries merge, they form a single object which undergoes a characteristic vibrational stage until it relaxes to a final stationary state. This state is approached via the ringdown phase. Due to its simplicity, it can be analyzed with perturbative schemes, by assuming that the background spacetime is slightly perturbed by very small fluctuations. In this limit, the Einstein field equations reduce to a master wave equation with an effective potential~\cite{Regge:1957td, Zerilli:1970se, 1974ApJ...193..443T, Chandrasekhar:1985kt} describing the radiative degrees of freedom of the gravitational field. The resulting equation can then be integrated in the time domain via scattering of waves off the BH effective potential~\cite{Vishveshwara:1970zz} or in the frequency domain to reveal the frequency content of the ringdown signal~\cite{Chandrasekhar:1975zza, Leaver:1986gd}. In both cases, the dynamics of the signal is sufficiently-described by a superposition of exponentially damped sinusoids. The oscillation frequency and decay timescale of the ringdown is associated with the real and imaginary parts of the QNM frequencies, which contain information regarding the underlying geometry of the remnant~\cite{Kokkotas:1999bd,Berti:2009kk,Konoplya:2011qq}. The QNM spectrum of BHs in GR has been extensively analyzed with various analytic and numerical methods~\cite{Pani:2013pma}, though the spectral content of ringdown signals generated by compact binaries is less understood. Even so, there are important indications that several BH oscillation harmonics, including the fundamental mode and higher overtones, as well as various multipolar components, are crucial to fully understand the GW ringdown~\cite{Berti:2005ys,Giesler:2019uxc}. 

\begin{figure}[t]\centering
	\includegraphics[scale=0.315]{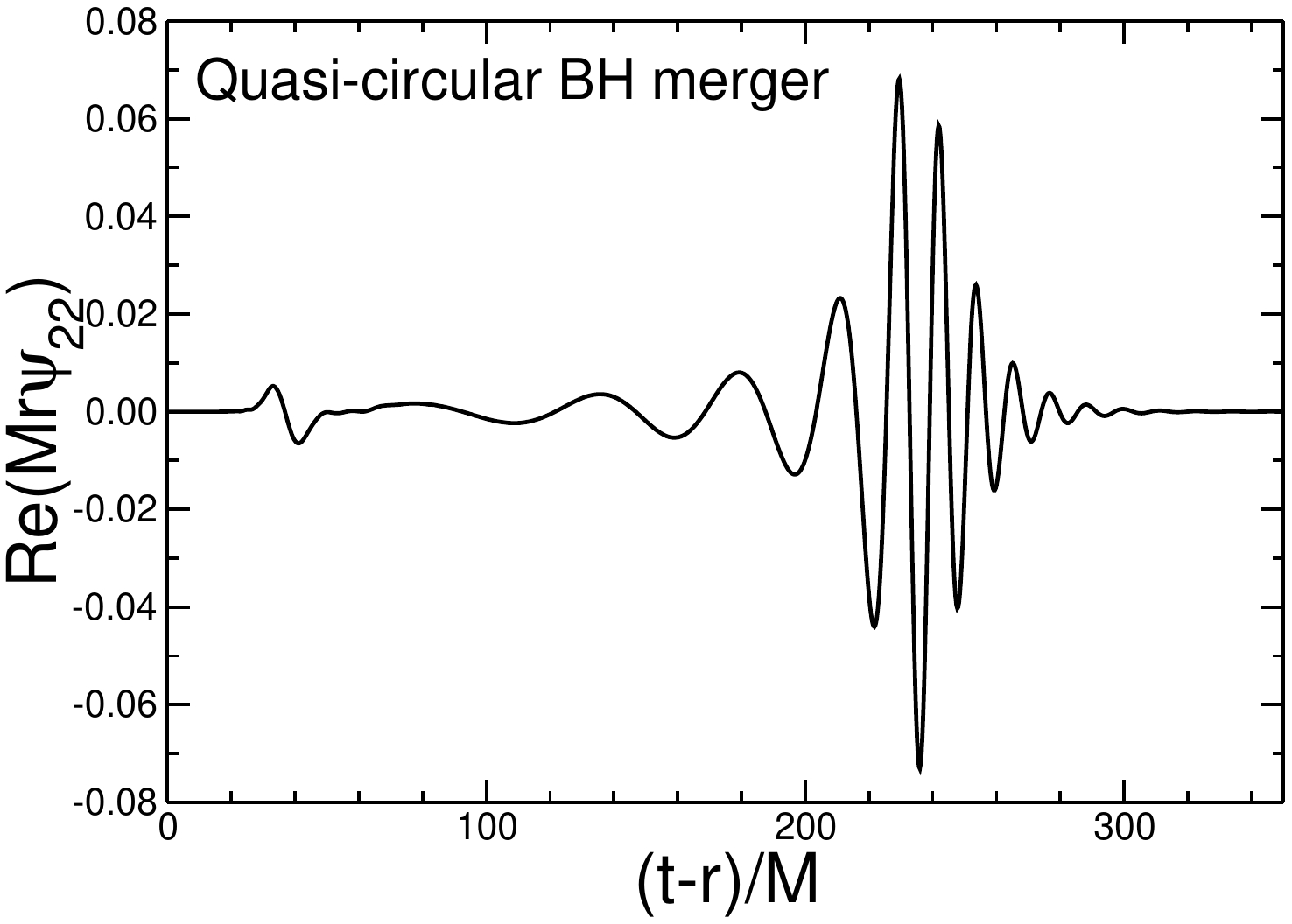}\hspace{-1cm}
	\includegraphics[scale=0.315]{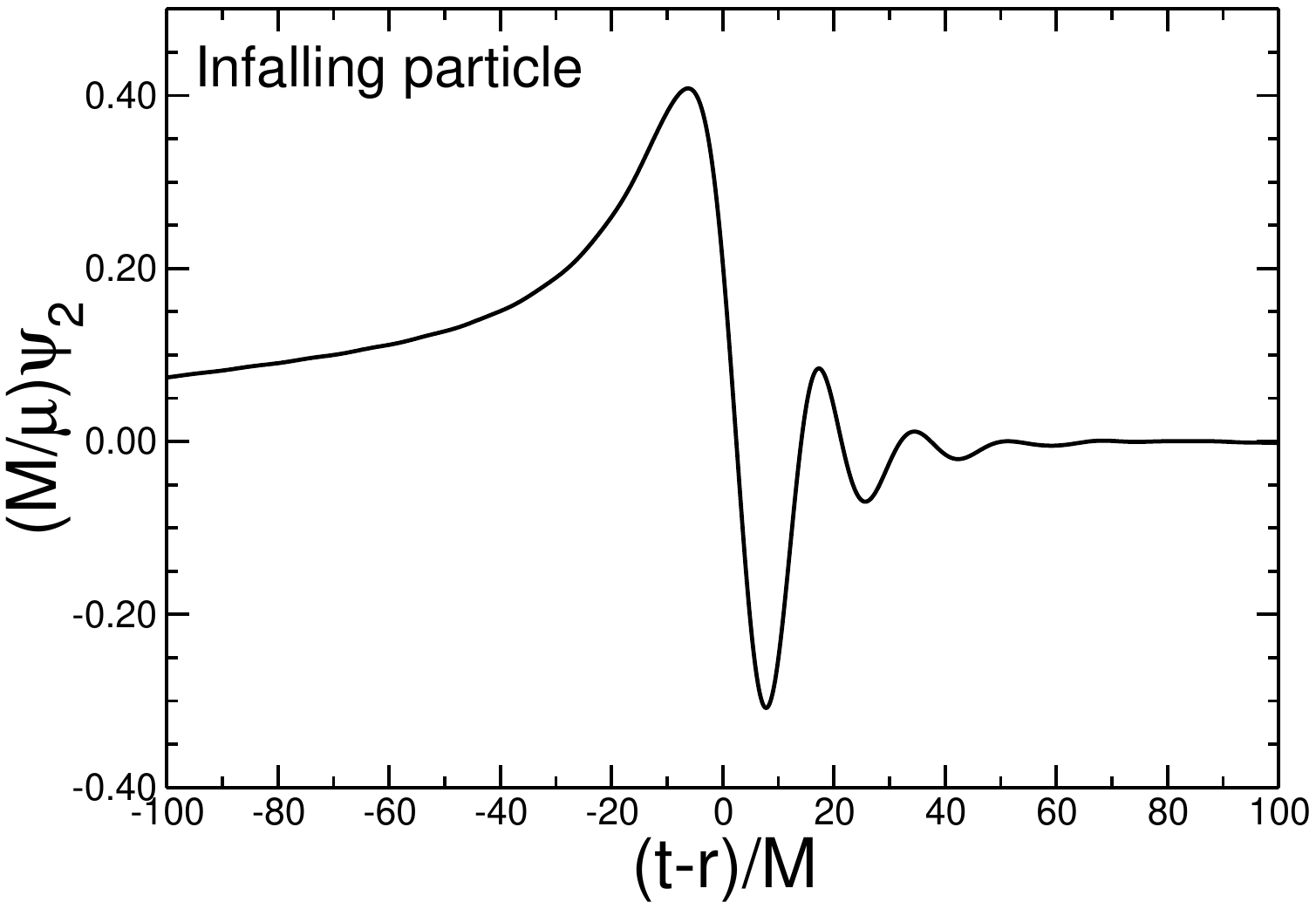}
	\caption{Left: The waveform produces by a quasi-circular BH binary merger, where $M$ is the total mass of the binary and $\Psi_{22}$ is $\ell=m=2$ multipolar component of the Weyl scalar $\Psi_4$. Right: The ringdown signal produced by an infalling particle with $\mu$ its mass onto a Schwarzschild BH with mass $M$, where $\Psi_2$ is the Zerilli function for $\ell=2$. In both cases, $r$ is the waveform extraction radius. Notice the similarities in the ringdown stage on both cases demonstrated. The figure is an adaptation from~\cite{Berti:2009kk}.}\label{ringdown}
\end{figure}

In what follows we adopt a static and spherically-symmetric background of the form
\begin{equation}\label{line_element}
	ds^2 = -f(r) dt^2 + f(r)^{-1} dr^2 + r^2 \left({d \theta^2} + \sin^2\theta d\varphi^2\right),
\end{equation}
where $f(r)$ is a lapse function that describes the spacetime background through its externally-observable parameters. The propagation of linear perturbations $\Psi$ on Eq.~\eqref{line_element} is described by the wave-like equation
\begin{equation}\label{wave_equation_time}
	\left(\frac{\partial^2}{\partial t^2} -\frac{\partial^2}{\partial r_*^2} + V \right)\Psi(t,r) = 0,\,\,\,\, V=f(r)\left(\frac{\ell(\ell+1)}{r^2}+(1-s^2)\frac{f^\prime(r)}{r}\right), 
\end{equation}
where $dr/dr_*=f(r)$ is the tortoise coordinate, $\ell$ is the angular number that appears from the decomposition of the perturbation onto spherical harmonics~\cite{Berti:2009kk}, $s=0, \pm1, \pm2$ correspond to scalar, electromagnetic and axial gravitational perturbations, respectively, and prime denotes radial differentiation. Equation~\eqref{wave_equation_time} can be evolved in the time domain~\cite{Gundlach:1993tp} in order to obtain the corresponding ringdown signal of perturbations on the fixed background~\eqref{line_element}. Figure~\ref{ringdown} represents such a scenario where the background is a Schwarzschild BH (with $f(r) = 1 -2 M/r$, where $M$ is the BH mass), together with a quasi-circular BH merger that produces a similar ringdown.
On the other hand, the time-domain perturbation evolution equation~\eqref{wave_equation_time} can be Fourier-transformed onto a time-independent Schr\"odinger-like equation in the frequency domain
\begin{equation}\label{wave_equation_frequency}
	\frac{\partial^2\Psi(r)}{\partial r_*^2} + \left(\omega^2-V\right)\Psi(r) = 0.
\end{equation}

\begin{figure}[h]\centering
	\includegraphics[scale=0.31]{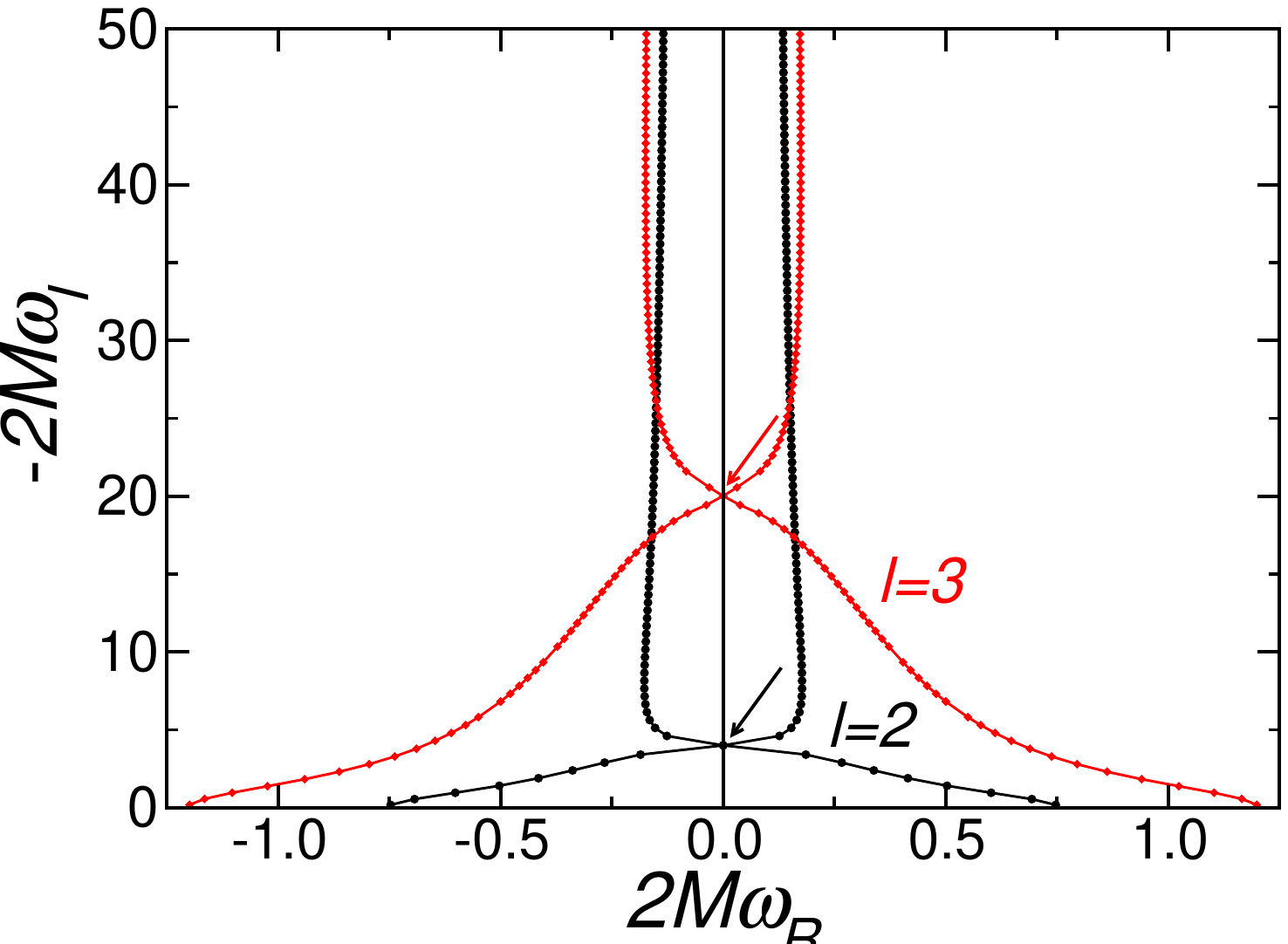}\hspace{-0.6cm}
	\includegraphics[scale=0.31]{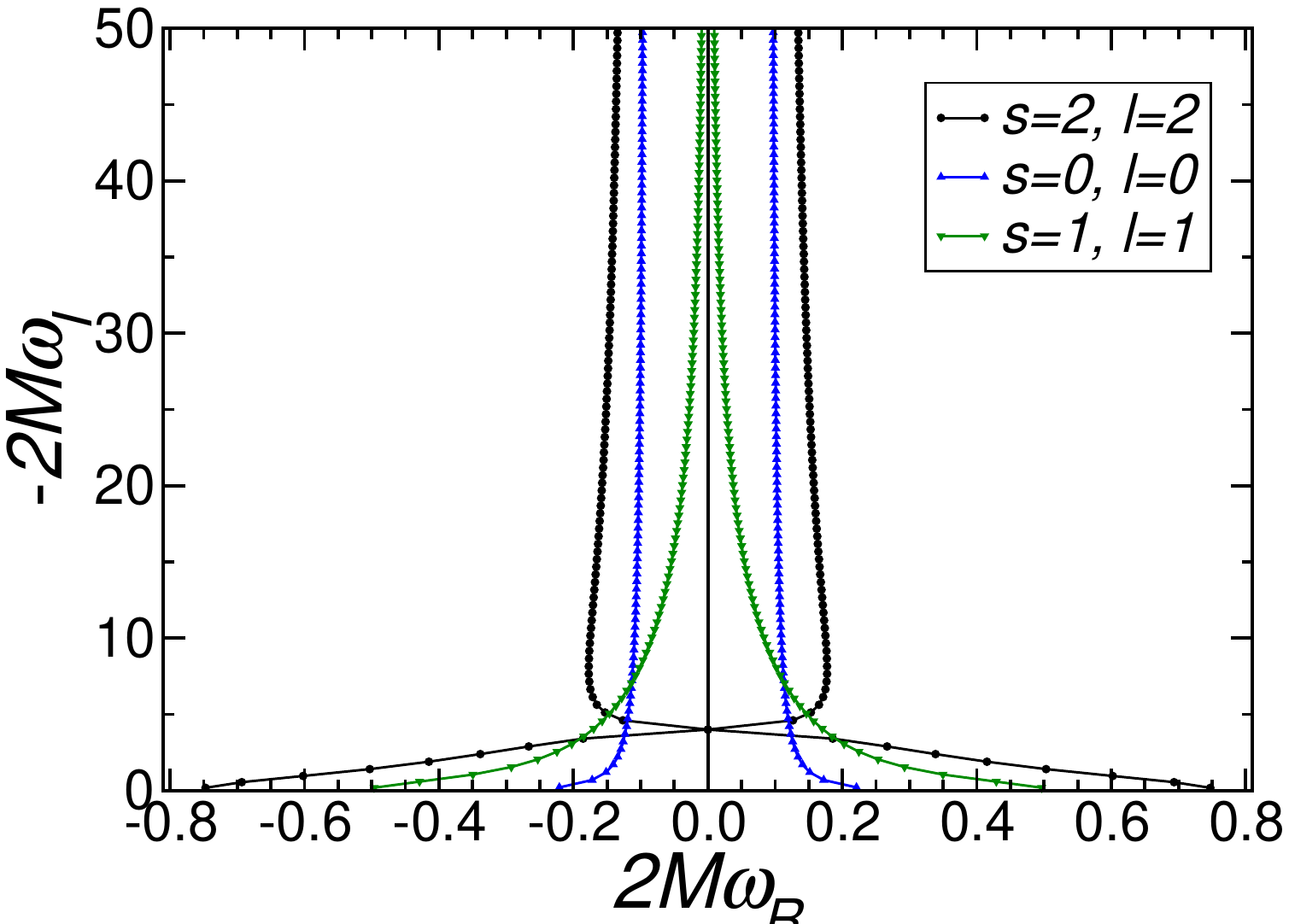}
	\caption{Left: Real $\omega_R$ and imaginary $\omega_I$ parts of axial gravitational QNMs of a Schwarzschild BH with mass $M$ and varying angular number $\ell$. Right: Real $\omega_R$ and imaginary $\omega_I$ parts of scalar ($s=\ell=0$), electromagnetic ($s=\ell=1$) and axial gravitational ($s=\ell=2$) QNMs of a Schwarzschild BH with mass $M$. The figure is taken from~\cite{Berti:2009kk}.}\label{QNMs}
\end{figure}

To obtain the QNM spectrum~\cite{Kokkotas:1999bd,Berti:2009kk,Konoplya:2011qq}, we apply appropriate boundary conditions for asymptotically-flat BHs, such that at the event horizon $r=r_h$ we only accept purely ingoing waves while at infinity we impose purely outgoing waves, i.e.
\begin{equation}
	\Psi \rightarrow
	\left\{
	\begin{array}{lcl}
		e^{-i\omega r_*}, & \mbox{  } & r_* \rightarrow - \infty \,\,\,\,\,\,(r\rightarrow r_h) 
		\\
		&
		&
		\\
		e^{i\omega r_*}, & \mbox{ } & r_* \rightarrow \infty \,\,\,\,\,\,\,\,\,\,\,(r\rightarrow \infty)
	\end{array}
	\right.
\end{equation}
A typical illustration of the QNM spectrum of Schwarzschild BHs is shown in Fig.~\ref{QNMs}.

\subsection{Quasinormal mode detection status}

The status of the fundamental mode detection is well-established since the first GW detection, GW150914~\cite{LIGOScientific:2016aoc}, though it is insufficient to recover accurately the mass and spin of the final remnant and perform validity tests of GR among other tests regarding BH environments, quantum effects and the existence of exotic compact objects that can mimick BHs in particular scenarios. The status of QNM overtones and if in fact we have managed to detect them in the loudest event to date, is currently under scrutiny, even though there was a strong certainty on their detection~\cite{Giesler:2019uxc,Isi:2019aib,Isi:2021iql,Isi:2022mhy}. 

In fact, the overtone status has been recently challenged regarding its robustness. In Ref. \cite{Cotesta:2022pci} it was pointed out that the uncertainty in the measured start time cannot be ignored, leading to a significant decrease in the evidence for overtones (see Ref. \cite{Carullo:2019flw} for earlier investigations). Additionally, GW150914-like injections in neighboring segments of the real detector noise at high sampling-rate have shown that noise can induce artificial evidence for overtones, indicating how overtone detection is quite sensitive against data-analysis choices. Such specific choices were the source of apparent initial disagreement~\cite{Cotesta:2022pci,Isi:2022mhy,Carullo_new}. When the time uncertainty is included, and identical settings employed, all analyses consistently point to a detection significance slightly below 2$\sigma$, as independently obtained through both time-domain~\cite{Isi:2022mhy,Carullo_new} and frequency-domain~\cite{Finch:2022ynt,Crisostomi:2023tle} methods. Furthermore, nonlinearities have been recently found at the early stage of BH ringdowns \cite{Cheung:2022rbm,Mitman:2022qdl,Kehagias:2023ctr,Kehagias:2023mcl,Perrone:2023jzq,Bucciotti:2023ets}. Such effects are expected to strongly impact the above analyses and their interpretation, due to the fact that the fitting begins arbitrarily close to the peak of the merger, which is highly dynamical and sensitive to time offsets and initial transients \cite{Zhu:2023mzv}.

The above issues created turmoil in the BH spectroscopy program, though at least the nonlinearity investigations now agree. Yet, the detection status of overtones is still under discussion, but with time the program will resolve this issue as well. Nevertheless, this chapter will argue how spectral instabilities may add another thorn in ringdown physics and how this might affect BH spectroscopy. This is particularly important for the overtone discussion, since they are the ones mostly affected from spectral instabilities.

\section{Early evidence of a spectral instability}

The first evidence of a spectral instability in BH physics dates back to the 90s, when Nollert attempted to solve the non-completeness issue of QNMs~\cite{Nollert:1996rf}. Due to the backscattering of perturbations at infinity, which lead to a late-time polynomial tail after the exponential ringdown~\cite{Gundlach:1993tp} and a branch cut in the complex plane of the retarded Green function of the wave equation~\cite{Leaver86c,Casals:2012ng} (this does not occur on non-asymtotically flat spacetimes), one cannot built a complete set of QNMs that describe the ringdown signal as a whole. Nollert's experiment consisted of a large number of step barriers in order to approximate and ultimately replace the continuous effective potential of gravitational perturbations in Schwarzschild BHs. His task was successful; the QNM spectrum of the step barrier potential could form a complete set though with a very strong drawback. The new \emph{perturbed spectrum} was entirely different from that of the \emph{unperturbed spectrum} (with continuous potential) of the BH~\cite{Nollert:1996rf}. Nevertheless, he later found that regardless that the perturbed QNMs were significantly different, the ringdown signal still behaved as if the unperturbed fundamental mode was present, and only at late times the new perturbed modes would appear~\cite{Nollert:1998ys}.

\begin{figure}[t]
	\includegraphics[scale=0.35]{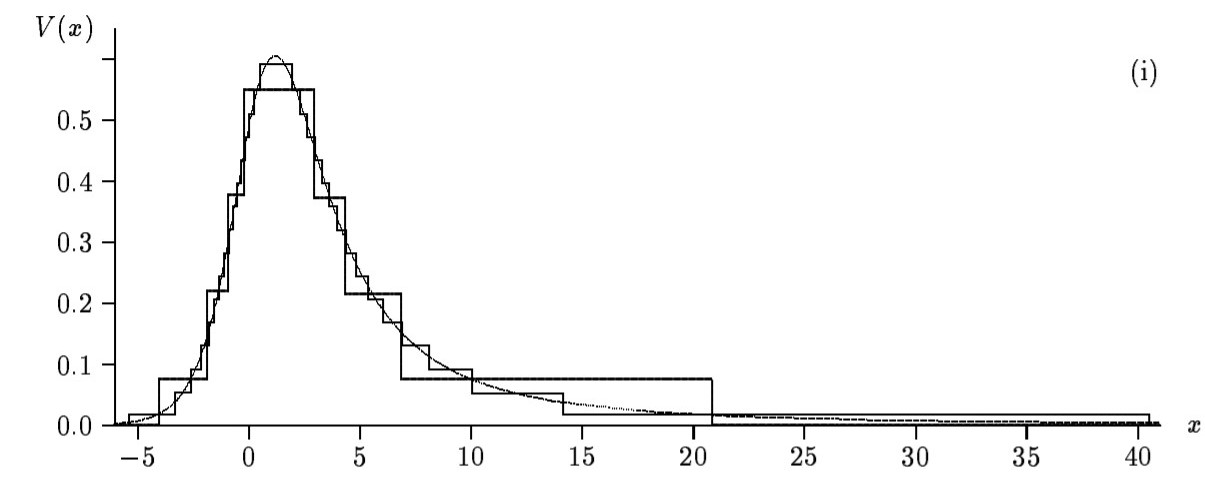}
	\includegraphics[scale=0.35]{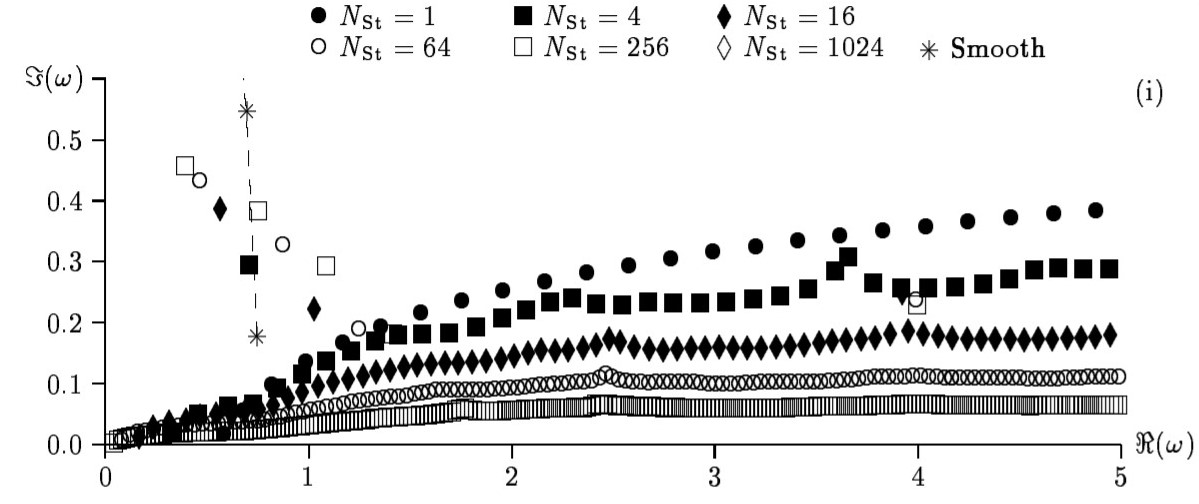}
	\caption{Left: The continuous BH potential from~\eqref{wave_equation_time}, with $\ell=s=2$, shown with a black line, and the various choices of step barriers to approximate it. Right: The corresponding spectrum resulting from the approximated potential, where $N_\text{St}$ is the number of steps in the potential. For reference, the modes shown with asterisks correspond to the continuous potential and their trend is shown with a dashed line. The figure has been adapted from~\cite{Nollert:1996rf}.}
	\label{Nollert_branch}
\end{figure}

In Fig.~\ref{Nollert_branch}, the methodology of approximating the potential with steps is demonstrated as well as the respective spectra that depend on the number of steps used. As already discussed, this was an unexpected result at the time, since the more steps were used, the more the perturbed QNMs were branching out and showing similarities with the w-modes of neutron stars~\cite{Kokkotas:1992xak}. The reason behind the spectra changing so significantly is obvious. When utilizing steps to approximate a potential, you cut out the asymptotic behavior where the first step appears. In all cases and no matter the number of steps used, besides infinity, the branch cut in the Green function is disregarded from the problem, which is the sole reason why the polynomial tail appears. One can then build a complete base of QNMs that describes the ringdown signal, since the branch cut is absent, but since the asymptotic behavior is altered abruptly at some distance $x$ in Fig.~\ref{Nollert_branch}, or equivalently at some $r_*$ in Eq.~\eqref{wave_equation_frequency}, it is natural that the spectrum will change, since the boundary conditions are intrinsically altered. This work paved the way to further understand the connection between QNMs of BHs and their role in ringdown signals and eventually led to the realization that BH spectroscopy may suffer from deep mathematical issues; in this case, spectral instabilities. For a closer look into this, see also~\cite{Daghigh:2020jyk}.

\section{The need for non-modal tools: the pseudospectrum}

To amply exploit the abilities of BH spectroscopy, one must better understand the relative excitation of each QNM~\cite{Nollert:1998ys} and the possible instability of the spectrum itself under small perturbations of the scattering potential~\cite{Nollert:1996rf}, as demonstrated in the previous section. Because astrophysical BHs are not isolated, but rather are permeated by highly dynamic environments, this last question is of paramount importance. For example, if the BH's surrounding is \emph{perturbed} in a small way by an accretion disk or matter shell, then the QNM spectrum may surprisingly change by a unequally large amount~\cite{Nollert:1996rf,Leung:1997was,Bamber:2021knr,Cheung:2021bol}. This migratory behavior of QNMs may have important implications for interpreting GW observations.

For selfadjoint operators, the spectral theorem underlies the notion of normal modes of vibration (which provide an orthonormal basis) and guarantees the stability of the eigenvalues under fluctuations of the eigenvalue operator. In other words, a small-scale perturbation to the operator leads to a new spectrum that migrates in the complex plane within a region of radius comparable to the scale of the perturbation, desgnating \emph{spectral stability}. In stark contrast, the lack of a spectral theorem in the non-selfadjoint case entails, in general, the loss of completeness in the set of eigenfunctions as well as their orthogonality. Thus, the eigenvalues may show strong sensitivity to small-scale environmental perturbations of the operator, as Nollert first found~\cite{Nollert:1996rf}. In these cases, the perturbed spectrum migrate to an extent that is orders of magnitude larger than the perturbation's scale, designating a \emph{spectral instability}. Operators that describe BH oscillations and QNMs fall into the second category.

\subsection{The mathematical notion of the pseudospectrum}

It turns out that there is a powerful mathematical notion called the pseudospectrum~\cite{Trefethen2000}, which is able to capture the extent to which systems controlled by non-selfadjoint operators exhibit spectral instabilities. The pseudospectrum is a continuous superset of the spectrum and can be formally defined through the \emph{resolvent operator} which portrays a `topographical map' of spectral migration under generic perturbations of order $\epsilon$ to the underlying eigenvalue operator. Formally, the pseudospectrum is defined as follows. Given $\epsilon>0$, the $\epsilon$-pseudospectrum $\sigma^\epsilon(A)$ of an operator $A$ can be characterized as
\begin{align}\label{def_pert}
	\sigma^\epsilon(A)&=\{z\in\mathbb{C}, \exists \; \delta A\!\in\! M_n(\mathbb{C}), ||\delta A||<\epsilon: z\!\in\!\sigma(A+\delta A) \},\\
	\sigma^\epsilon(A)&=\{z\in\mathbb{C}:   ||R_A(z)||=||(z\mathbb{I}- A)^{-1}||>1/\epsilon\},\label{def_resolvent}
\end{align}
with $z$ a complex number, $\delta A$ a complex operator, $R_A(z)$ the resolvent operator of $A$ at the point $z$, $\mathbb{I}$ the identity operator and $\sigma(A)$ the spectrum of $A$. When $\epsilon\rightarrow 0$, then $\sigma^0(A)=\sigma(A)$.

\begin{figure}[b]\centering
	\includegraphics[scale=0.55]{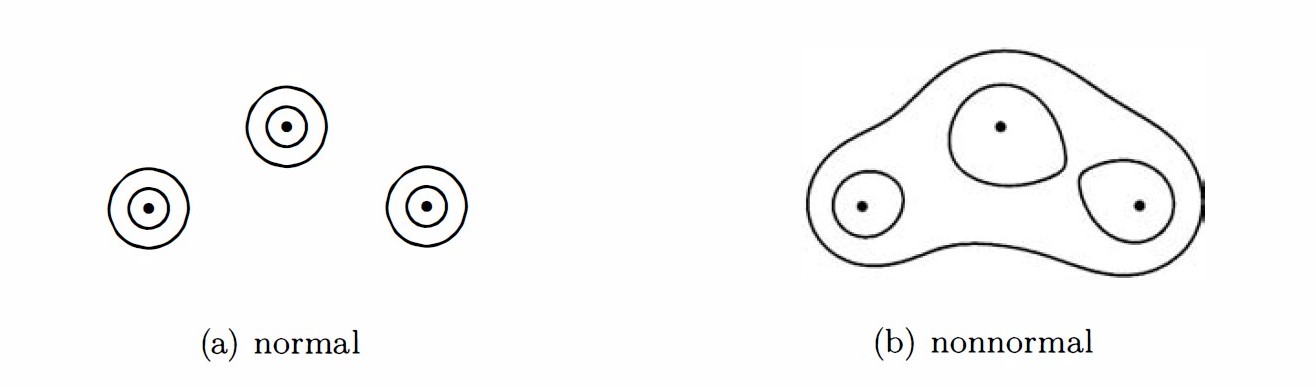}
	\caption{The geometry of pseudospectra of a self-adjoint (normal) (a) and a non-selfadjoint (non-normal) (b) operator. In each plot, the contours represent the boundary of $\sigma^\epsilon(A)$ for two values of $\epsilon$, while the eigenvalues are shown with black dots. The figure is adapted from~\cite{Trefethen:1993}.}
	\label{contours}
\end{figure}

\begin{figure}[t]\centering
	\includegraphics[scale=0.5]{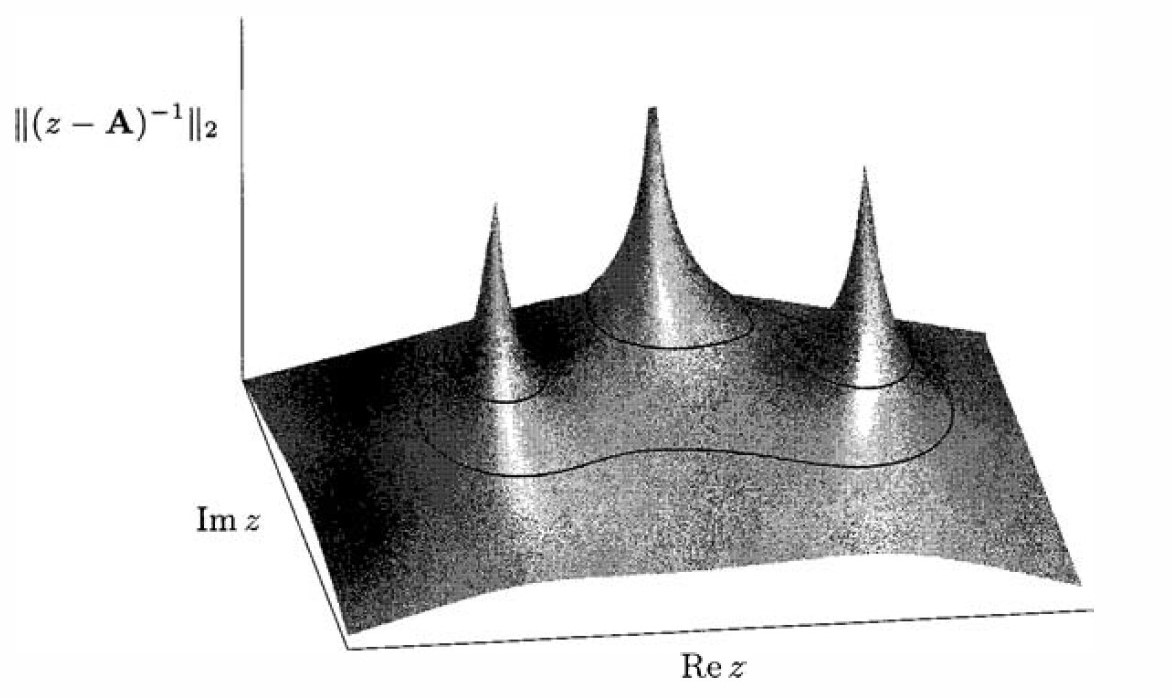}
	\caption{The resolvent norm as a function of $z$ in a specified region of the complex plane for case (b) in Fig.~\ref{contours}. The contours shown in black match those in Fig.~\ref{contours}, (b). The peaks correspond to eigenvalues of the assumed operator, where the norm of the resolvent diverges. The figure is adapted from~\cite{Trefethen:1993}.}
	\label{resolvent}
\end{figure}

Selfadjoint (normal) operators satisfy $\left[A,A^\dagger\right]=0$, therefore the spectral theorem of normal operators ensures that eigenfunctions form an orthonormal basis and that the perturbed eigenvalues of $\sigma(A+\delta A)$, in accord to Eq.~\eqref{def_pert}, are spectrally stable, i.e. the perturbed spectra will reside in concentric circles of radius $\epsilon$. Spectral stability can also be understood in a more robust way by Eq.~\eqref{def_resolvent}. Assume a point in the complex plane $z$. Asking if it is an eigenvalue is not a robust question (numerically) since the equation $\det|z\mathbb{I}- A|=0$ has to be satisfied exactly, which needs fine-tuning or extremely accurate predictions. Nevertheless, selfadjoint operators abide with an eigenvalue analysis since it gives good predictions regarding stability. On the contrary, non-selfadjoint operators (non-normal) lack a spectral theorem and lead to weak control of eigenfunction completeness and stability. Therefore, a more well-posed question to ask is that of Eq.~\eqref{def_resolvent}, i.e. how close to an eigenvalue a point of the complex plane is through the resolvent operator. Hence, the pseudospectrum, defined either by~\eqref{def_pert} or~\eqref{def_resolvent}, can answer questions regarding eigenvalues, their spectral stability and associated eigenfunction completeness. Figure~\ref{contours} portrays a schematic representation of hypothetical normal and non-normal operators spectra and $\epsilon$-pseudospectra. The differences between the structure of these operators is quite obvious. The normal operator forms concentric pseudospectra contours around the eigenvalues, while the non-normal operator contours extend further from the eigenvalues and do not form concentric circles with specified radii. The three-dimensional Fig.~\ref{resolvent} for the non-normal case demonstrates the topographic structure of the pseudospectrum through the resolvent's norm in the complex $z$-plane.

As demonstrated, strong non-normality, occuring in quite generic settings, leads to a severely uncontrolled spectrum under perturbations of the governing operator, thus by using the pseudospectrum in BH physics we can formally capture the potential sensitivity of the QNMs to small perturbations at the level of the non-perturbed operator that described BH QNMs, without the need of systematically introducing perturbations to the operator, as suggested by Eq.~\eqref{def_pert}. As a side note, we need to emphasize that the definition of the pseudospectrum, and thus any statement on spectral stability, depends delicately on the choice of the norm used, which directly connects with a scalar product. The scalar product fixes the norm that quantifies the extent of large and small perturbations $||\delta A||$. Physically, one of the logical choices for a norm is the one that captures the system's energy, since it conventionally encodes the size of the physical perturbation with respect to the physics of the problem. In what follows, the so-called \emph{energy norm}, or energy scalar product, which is systematically discussed in~\cite{Gasperin2021} and introduced decades ago in~\cite{Driscoll:1996}, will be used. We point the reader to~\cite{Gasperin2021} and in the Appendices of~\cite{Jaramillo2020} for a complete discussion of the energy norm as well as how it can be incorporated in the resolvent operator and eventual pseudospectrum computation.


Naturally, there is a variety of non-modal tools, such as the \emph{eigenvalue condition number}, the \emph{spectral, numerical, pseudospectral abscissa} and the \emph{Kreiss constant}, among others, that help to further understand the transient early-time behavior of the matrix evolution operator and the existence of pseudoresonances \cite{Trefethen:2005}. In this chapter, we will stay clear from these concepts and focus on the \emph{pseudospectrum of compact objects}, though we encourage the reader to take a close look at~\cite{Trefethen:2005,Jaramillo:2022kuv,alsheikh:tel-04116011} for further details.

\subsection{Hydrodynamics and the pseudospectrum}\label{hydro}

Mathematical tools that do not depend on the calculation of eigenvalues are not new in the literature of science. The non-modal notion of the pseudospectrum has been used for decades in issues such as quantum mechanics, where the introduction of non-selfadjoint operators in PT-symmetric quantum mechanics entails that the associated spectrum is insufficient to draw full, quantum-mechanically relevant conclusions~\cite{Krejcirik:2014kaa}. Another application, which is also extremely pertinent to spherically-symmetric BH perturbations, is on one-dimensional wave equations with dissipative boundary conditions which suffer from similar limitations in spectral predictions~\cite{Driscoll:1996}. 

Probably one of the most interesting application of the pseudospectrum and its potential to predict spectral instabilities and even nonlinear phenomena at the linear operator level occurred in the field of hydrodynamics. The linearized Navier-Stokes equations are typically used to theoretically understand if a laminar (stable) fluid flow will become turbulent (unstable) for a choice of parameters that characterize the initial fluid flow, such as its speed. Eigenvalue analysis predicted that there is a critical (or no) parameter of the fluid flow beyond which the instability occurs~\cite{orszag_kells_1980}. To the contrary, in the laboratory, transition to turbulence was observed for much smaller fluid parameters than those predicted by modal analyses~\cite{patel_head_1969,carlson_widnall_peeters_1982}. These anomalies between theory and experiment of subcritical transition to turbulence and their initial explanation have been recognized for decades as bad attributes of the linearization around the laminar flow.

\begin{figure}[h]\centering
	\includegraphics[scale=0.6]{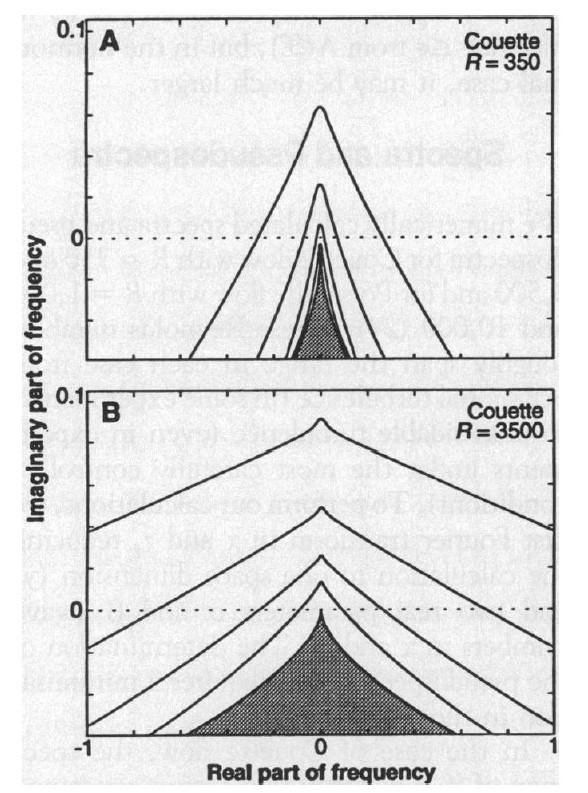}\hspace{0.2cm}
	\includegraphics[scale=0.6]{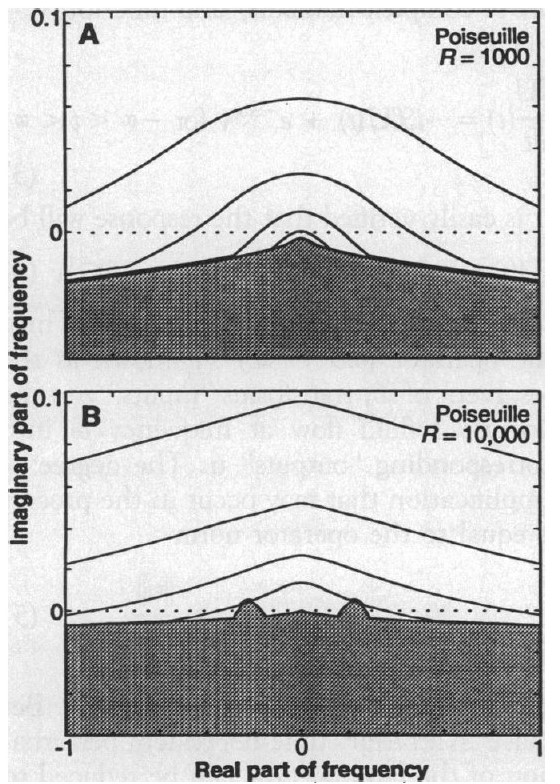}
	\caption{Left: Spectrum (gray area) and $\epsilon$-pseudospectra (solid curves) in the complex plane of the linearized Navier-Stokes evolution operator for Couette hydrodynamic flow with specific Reynolds number $R$ that is proportional to the speed of the flow. The pseudospectral contour lines from outer to inner correspond to $\epsilon=10^{-2},\, 10^{-2.5},\,10^{-3},\,10^{-3.5}$. The horizontal dashed lines designate the real axis, above which unstable modes lie. The pseudospectrum extends significantly on the upper half of the complex plane, in which by convention, unstable modes reside. Right: Same as Left but for Poiseuille hydrodynamic flow. For $R>5772$, two bumps of the eigenvalue area extend into the upper half of the complex plane. The figure is adapted from~\cite{Trefethen:1993}.}
	\label{hydro_pseudospectrum}
\end{figure} 

Later, it was understood that the failure of the eigenvalue analysis may be attributed to a misguided search for unstable eigenvalues of the linearized system. The pseudospectrum gave the answer to the aforementioned anomalies in a seminal work of Trefethen \emph{et al.}~\cite{Trefethen:1993}. The analysis performed in hydrodynamics and the Navier-Stokes equations~\cite{Trefethen:1993}, and later the books written regarding the pseudospectrum of various physical systems~\cite{Trefethen2000,Trefethen:2005}, revealed that the pseudospectra of the linearized problem implied that small perturbations to the smooth flow may be amplified by factors of order $10^5$ by a linear mechanism, even though all the eignevalues of the system decay monotonically. This occurs due to the non-orthogonality of the eingenfunctions and the non-normality of the underlying operator of the hydrodynamic system. The most notable results from~\cite{Trefethen:1993} are demonstrated in Fig.~\ref{hydro_pseudospectrum}, where the spectrum is shown with in the dark gray area. It is evident that the subcritical transition to turbulence is encoded in the pseudospectrum of the linearized Navier-Stokes equations, in total agreement with experiments, due to the protrusion of the pseudospectral contours in the upper half of the complex plane, where unstable modes reside. This is an indication that the tiniest perturbation to the governing operator of the linear system can destabilize the evolution and lead to turbulent flow, even though all eigenvalues lie on the lower half of the complex plane, which designates modal stability. For further indications regarding spectral instabilities and the need for non-modal tools, we refer the reader to~\cite{Trefethen:1993,Trefethen:2005}, though in what follows, we will focus solely on the pseudospectrum.

\subsection{Black holes and their pseudospectrum in a nutshell} 

The pseudospectrum was recently introduced in the context of BH physics to address BH QNM instabilities. The seminal analysis in~\cite{Jaramillo2020} indicates an overall spectral instability in Schwarzschild BH QNMs. Similar spectral instability have been found in subextremal and extremal Reissner-Nordstr\"om (RN) BH QNMs~\cite{Destounis:2021lum}, exotic horizonless objects \cite{Boyanov:2022ark}, as well as BH with de Sitter \cite{Sarkar:2023rhp} and anti-de Sitter asymptotics \cite{Arean:2023ejh}. The mathematical tools introduced in~\cite{Jaramillo2020} have the potential to be used in order to address many interesting problems in gravitational physics, such as testing GR with GWs. Such a test would require the ability to extract various QNMs from the ringdown of BH mergers. By detecting the fundamental mode, we can determine the mass and spin of the final BH. Nevertheless, if the overtones could be accurately extracted, they would provide a test of GR, since overtones differ a lot in different theories of gravity~\cite{Cardoso:2019mqo,McManus:2019ulj,Kimura:2020mrh,Volkel:2022aca,Volkel:2022khh,Franchini:2022axs}. It turns out that the predicted instability in these overtones may limit the abilities of BH spectroscopy~\cite{Jaramillo:2021tmt} though further investigations are needed to assess to what extent a BH spectral instability at the ringdown signal can be detected with LIGO and LISA. Since the pseudospectrum is hardwired to predict transient phenomena and may assess up to which extent a BH QNM instability can limit the abilities of BH spectroscopy~\cite{Jaramillo:2021tmt}, it is of high importance to further explore its applicability in the vast zoo of BH physics. In what follows, we present a plethora of recent findings that regard spectral stability of BH and exotic compact object (ECO) QNMs, their corresponding pseudospectra, a more modern analysis of spectral instability detection in ringdown signals in line with~\cite{Nollert:1998ys}, as well as future directions to be explored.

\subsection{A pseudospectrum calculator primer}\label{pseudo_calculation}

\begin{figure}[b!]\centering
	\includegraphics[scale=0.25]{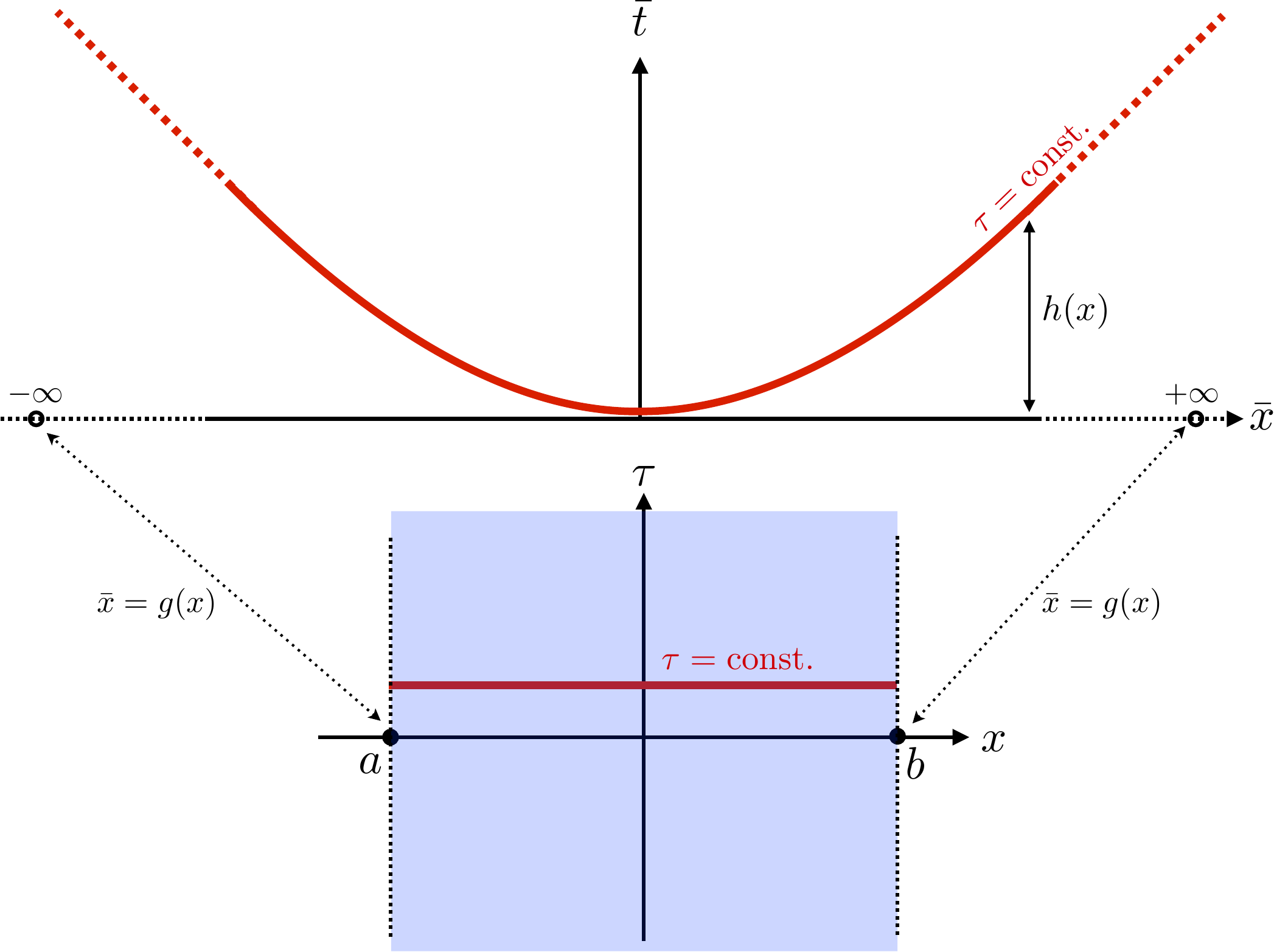}
	\caption{Schematic representation of the implementation of the hyperboloidal scheme, together with the spatial compactification. The height function $h(x)$ bends the Cauchy data upwards to penetrate the horizon and null infinity, while the function $g(x)$ compactifies the domain from $\bar{x}\in[-\infty,\infty]$ to $x\in[a,b]$. The figure is taken from~\cite{Jaramillo2020}.}
	\label{hyperboloidal}
\end{figure}

In order to numerically calculate pseudospectra of operators and matrices, and in particular those that describe the QNM problem of asymptotically flat BHs, it has been found in~\cite{Ansorg2016,PanossoMacedo:2018hab,PanossoMacedo:2020biw} that an appropriate coordinate transformation is necessary in order to build a matrix, that describes the QNM problem, and later discretize it. This transformation is known as the \emph{hyperboloidal scheme} to QNMs which extends the domain of $r_*\in\left[-\infty,\infty\right]$ and introduce in Eq.~\eqref{wave_equation_time} the dimensionless quantities
\begin{align}\label{rescale}
	\bar{t}=t/\lambda, \ \ \bar{x}=r_*/\lambda, \ \
	\tilde{V}_\ell=\lambda^2 V_\ell,
\end{align}
where $\lambda$ is an appropriate length scale to be defined according to the spacetime in study. We consider coordinates $(\tau,x)$ that implement the hyperboloidal foliation~\cite{Ansorg2016,Jaramillo2020}
\begin{align}\label{hyper_compact}
	\bar{t} = \tau - h(x),  \ \ 
	\bar{x} = g(x), 
\end{align}
where the height function $h(x)$ implements the hyperboloidal slicing by bending the Cauchy initial data to horizon and infinity penetrating data~\cite{Zenginoglu:2007jw,Zenginoglu:2011jz} (see Fig.~\ref{hyperboloidal} for a schematic representation) and the function $g(x)$ introduces a spatial compactification from $\bar{x}\in[-\infty,\infty]$ to $x\in[a,b]$.
The functions $h(x)$, $g(x)$ depend on the spacetime under study and are not fixed for all compact objects (see e.g. \cite{Jaramillo2020,Destounis:2020pjk,Boyanov:2022ark}).

Through Eqs.~\eqref{rescale} and~\eqref{hyper_compact}, Eq.~\eqref{wave_equation_time} becomes~\cite{Ansorg2016,PanossoMacedo:2018hab,Jaramillo2020,Destounis:2021lum}
\begin{equation}\label{hyperboloidal evolution equation}
	- \ddot \phi + L_1 \phi + L_2 \dot \phi = 0,
\end{equation}
where an overdot denotes differentiation with respect to $\tau$, and the differential operators are given by~\cite{Jaramillo2020}
\begin{align}\label{L1}
	L_1 = \frac{1}{w(x)}\big[\partial_x\left(p(x)\partial_x\right) - q_\ell(x)\big], \ \
	L_2 = \frac{1}{w(x)}\big[2\gamma(x)\partial_x + \partial_x\gamma(x)\big].
\end{align}
The height and compactification functions $h(x)$ and $g(x)$ enter the above operators via
\begin{equation}\label{new operators}
	w(x)=\frac{g'^2-h'^2}{|g'|}, \ \ \ \ p(x) = |g'|^{-1}, \ \ \ \
	\gamma(x)=\frac{h'}{|g'|}, \ \ \ \ q_\ell(x)= \lambda^2 |g'|\;V. 
\end{equation}
Equation~\eqref{hyperboloidal evolution equation} can be re-written as a matrix evolution problem through a first-order reduction in time by introducing $\psi=\dot \phi$. Equation~\eqref{hyperboloidal evolution equation} becomes
\begin{equation}\label{matrix evolution}
	\dot u=i L u, \quad 	L =\frac{1}{i}\!
	\left(
	\begin{array}{c  c}
		0 & 1 \\
		L_1 & L_2
	\end{array}
	\right), \quad u=\left(
	\begin{array}{c}\phi \\ \psi \end{array}\right),
\end{equation}
which has the eigenfunction solution
\begin{equation}
	\label{e:evolution_operator}
	u(\tau,x)=e^{iL\tau}u(0,x) \, ,
\end{equation}
in terms of the evolution operator $e^{iL\tau}$. By further performing a harmonic decomposition $u(\tau,x)\sim u(x)e^{i\omega\tau}$ in Eq.~\eqref{matrix evolution} we arrive at the eigenvalue equation
\begin{equation}\label{eigenvalue problem}
	L\, u_{n\ell}=\omega_{n\ell}\, u_{n\ell}, 
\end{equation}
where $\omega_{n\ell}$ is an infinite set of eigenvalues of the operator $L$, with $n\geq 0$ labelling the overtone number. The calculation of QNMs through the hyperboloidal framework ultimately reduces to the eigenvalue problem of the operator $L$, which, in turn, contains information concerning the boundary conditions of the spacetime metric at the geometric level. It is important to note that the blocks $L_1$ and $L_2$ multiply $\phi$ and $\dot{\phi}$, respectively. Hence, we can intuitively understand that $L_1$ contains terms associated with the effective potential (it includes $q_\ell\sim V_\ell$) and $L_2$ incorporates the boundary conditions at the event horizon and infinity (dissipative effects) due to its proportionality with the height function that applies the hyperboloidal slicing. This will prove important later on and serve as a sanity check for using the energy norm~\cite{Jaramillo2020,Gasperin2021}.

To bring the aforementioned system in a numerical format, we discretize the compactified domain of the operator $L$, $x\in\left[0,1\right]$, with $N+1$ Chebyshev grid points, while to discretize the differential operators we apply the Chebyshev differentiation matrices~\cite{Trefethen2000,Trefethen:2005}. Once $L$ is discrete, the BH QNMs follow from the eigenvalues of the matrix. For pseudospectra, one implements the Chebyshev-discretized version of the energy norm to calculate the adjoint $L^\dagger$, which is exhaustively detailed in~\cite{Gasperin2021}, the Appendices of~\cite{Jaramillo2020} and in~\cite{Destounis:2021lum}. The $\epsilon$-pseudospectrum $\sigma^\epsilon(L)$ in the energy norm is given by~\cite{Jaramillo2020}
\begin{equation}
	\label{pseudospectra energy norm}
	\sigma^\epsilon(L) = \{\omega\in\mathbb{C}: s^\mathrm{min}(\omega \mathbb{I}- L)<\epsilon\}, 
\end{equation}
where $s^\mathrm{min}$ is the minimum of the generalized singular value decomposition, which incorporates the adjoint in the energy scalar product
\begin{equation}
	\label{svd definition}
	s^\mathrm{min}(\mathcal{M}) = \min \{\sqrt{\omega}:  \omega\in \sigma(\mathcal{M}^\dagger \mathcal{M}) \}, \ \ \ \ \ \mathcal{M}=\omega \mathbb{I}- L.
\end{equation}

\section{Pseudospectrum of black holes}

In the following section, we present the most interesting results obtained in the seminal work of Ref.~\cite{Jaramillo2020} when calculating the pseudospectrum of a toy model BH, which includes a P\"oschl-Teller potential and the pseudospectrum of Schwarzschild BH QNMs, as well as the pseudospectrum and its rich phenomenology in RN BH QNMs~\cite{Destounis:2021lum}.

\subsection{P\"oschl-Teller potential: a toy model}

In this toy model, the Schwarzschild effective potential for QNMs in Eq.~\eqref{wave_equation_time} is replaced by the P\"oschl-Teller function
\begin{equation}\label{PT_potential}
	V(\bar{x})=V_0/\cosh^2(\bar{x}), \,\,\,\,\, \bar{x}\in(-\infty,\infty),
\end{equation}
which is typically used in QNM studies~\cite{Ferrari:1984,Beyer:1998nu}, since for an appropriate choice of $V_0$, one can reproduce the fundamental axial gravitational QNM of a Schwarzschild BH with good accuracy. Moreover, the hyperboloidal scheme has already been tested in this setup~\cite{Bizon:2020qnd}. A key attribute of the P\"oschl-Teller potential is that it decays exponentially on both boundaries, thus does not include a branch cut in the complex plane of the Green's function, and linear field fluctuations decay exponentially at infinity.

\subsubsection{Pseudospectrum: spectral stability and instability}

By applying the methods outlined in section~\ref{pseudo_calculation}, we can bring the wave equation, with $V_\ell$ as in Eq.~\eqref{PT_potential}, in a matrix evolution problem form, that allows us to find the spectrum and pseudospectrum (see~\cite{Jaramillo2020} for a detailed description of the steps involved). It has been noted that the block matrix $L_2$ contains the dissipative boundary conditions needed for QNMs. With the P\"oschl-Teller potential, one can perform the following validity check outlined in~\cite{Jaramillo2020}: by setting $L_2=0$ and keeping $L_1$ as is, one obtains a selfadjoint operator, with eigenvalues that are normal modes. Such an operator is relevant by itself, since it corresponds to the azimuthal mode $m = 0$ of a wave propagating on a sphere with a constant unit potential, which is a conservative, rather than a dissipative, system. The eigenfunctions are then the Legendre polynomials $\phi_n(x) = P_n(x)$, with real eigenvalues $\omega^\pm_n=\pm\sqrt{1+\ell(\ell+1)}$. This provides a robust test case for the norm chosen to calculate pseudospectra and we should expect the existence of a spectral theorem and a flat pseudospectrum with concentric circles forming around the normal modes of the system.

\begin{figure}[t]\centering
	\includegraphics[scale=0.45]{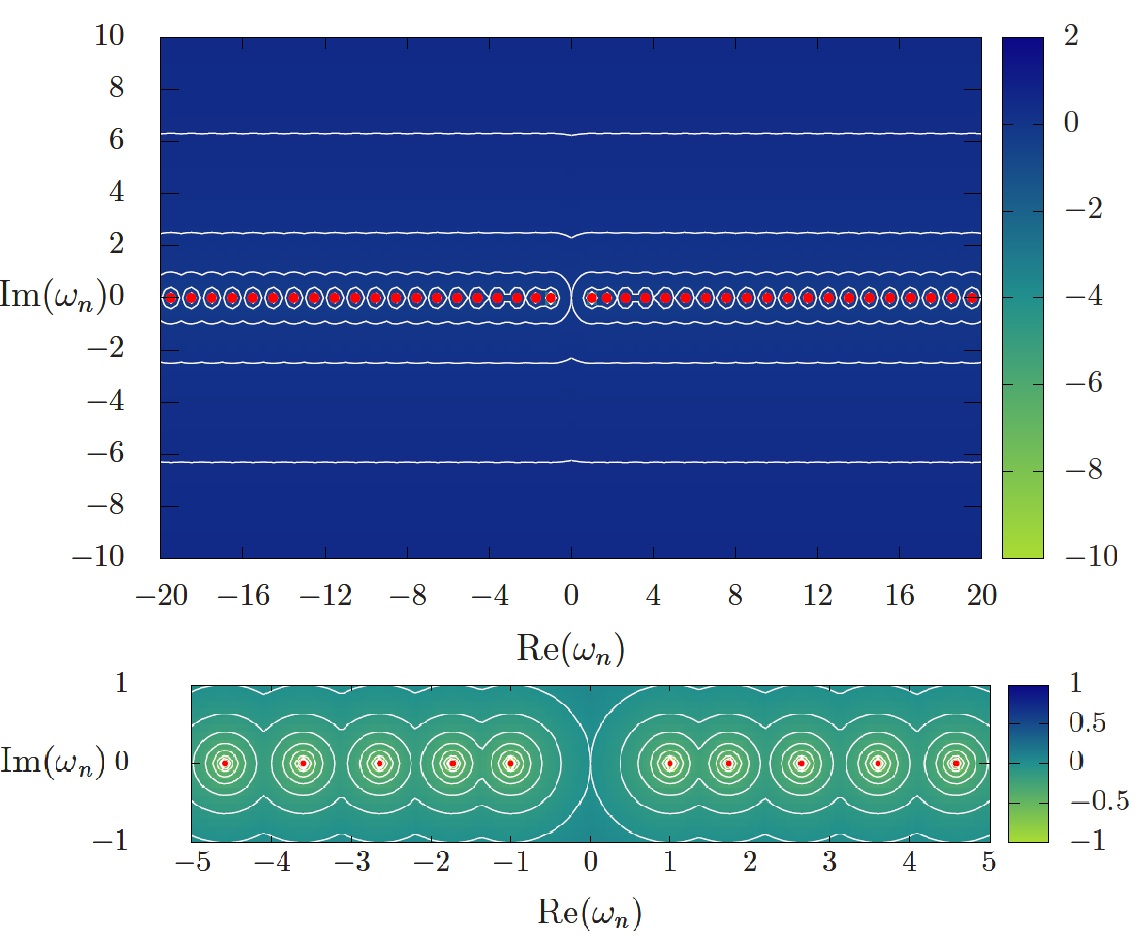}
	\caption{Pseudospectrum of a normal operator ($L_2=0$) with a P\"oschl-Teller potential with number of grid points $N=100$. The white lines correspond to different $\epsilon$ pseudospectral contour lines which become more negative close to the eigenvalues (red dots). The colorbar on the right is in logarithmic scale meaning that a value of, e.g. $-4$ corresponds to $\log[\sigma^{-4}(L)]$ or simply a perturbation in $L$ with energy norm $10^{-4}$. The figures have been adapted from~\cite{Jaramillo2020}.}
	\label{spectral stability}
\end{figure}

Figure~\ref{spectral stability} demonstrates exactly how a selfadjoint operator's pseudospectrum should look like. The absence of dissipative effects, such as the boundary conditions for QNMs, force the pseudospectrum to form concentric circles around the eigenvalues of the conservative system, where the circles have radii equal to the order of perturbation in the energy norm, i.e. $\epsilon$, as better shown in the bottom panel of Fig.~\ref{spectral stability}. The rest of the complex plane (shown in the upper panel) appears flat, i.e. having the same dark blue color everywhere and can only be accessed by the perturbed normal modes with very large-scale perturbations. Therefore, Fig.~\ref{spectral stability} represents how spectral stability manifests in conservative systems and provides a reassuring test of the whole numerical scheme introduced in~\cite{Jaramillo2020} regarding the use of the energy norm.

\begin{figure}[t]\centering
	\includegraphics[scale=0.65]{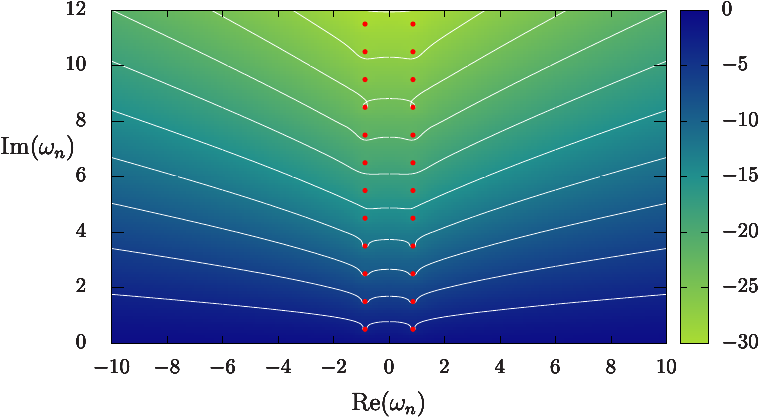}\hspace{0.1cm}
	\includegraphics[scale=0.32]{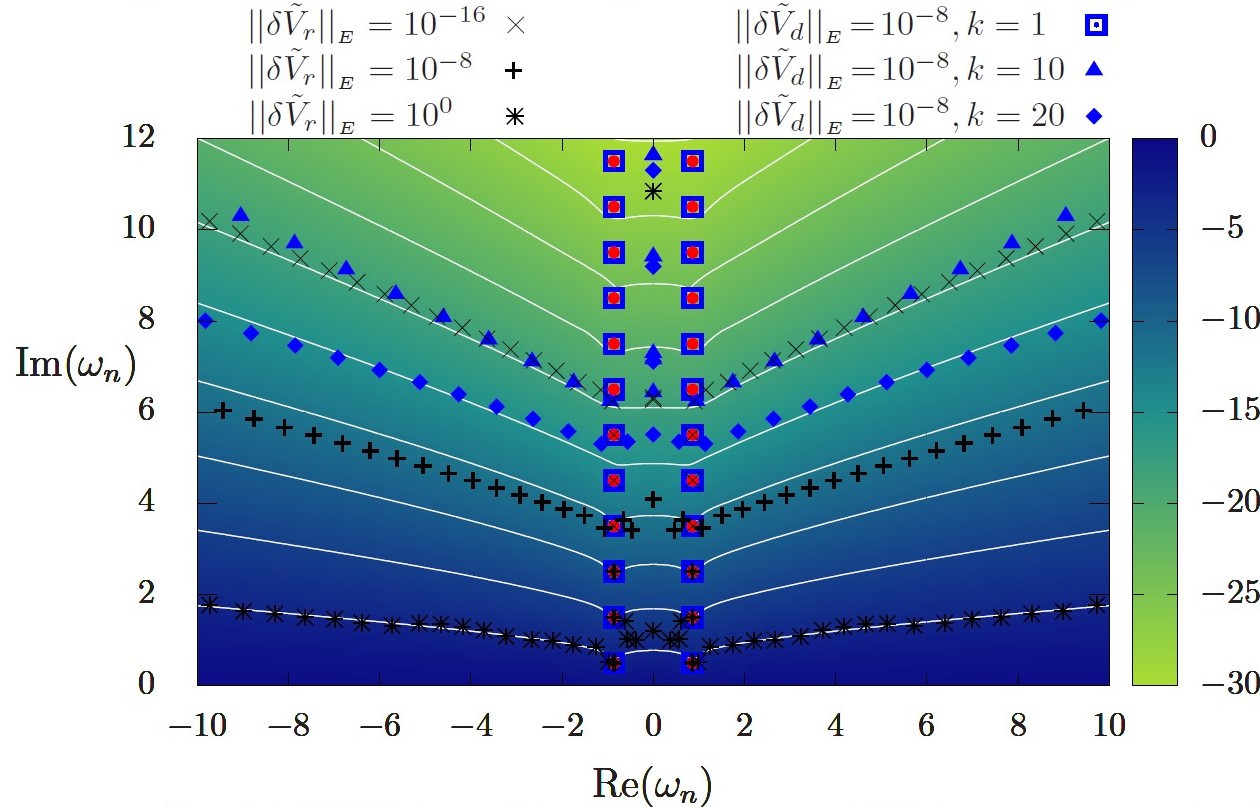}
	\caption{Left: Pseudospectrum of the wave equation~\eqref{wave_equation_time} with a P\"oschl-Teller potential and $N=100$ grid points. The white lines correspond to different $\epsilon$ pseudospectral contour lines, which become more negative close to the eigenvalues (red dots). The colorbar on the right is in logarithmic scale, meaning that a value of, e.g. $-15$ corresponds to $\log[\sigma^{-15}(L)]$ or simply a perturbation in $L$ with energy norm $10^{-15}$. Right: The pseudospectrum from Left with over-plotted perturbed QNMs resulting from random $\delta V_r$ and deterministic $\delta V_d$ perturbations to the potential. The perturbations are normalized such that in the energy norm they acquire the specified normalized value shown above the right panel. The figures have been adapted from~\cite{Jaramillo2020}.}
	\label{spectral instability}
\end{figure}

Of course, our interest towards BH QNMs and their spectral stability requires that we apply the proper ingoing boundary conditions at the (would-be) event horizon and outgoing at infinity, as the hyperboloidal approach already implements, meaning that $L_2\neq 0$. On the left panel of Fig.~\ref{spectral instability}, the spectral instability of the eigenvalues is obvious with pseudospectral contours extending logarithmically to asymptotic regimes that are disproportionally larger than those assumed in the operator $L$, through the resolvent norm. Therefore, as a rule of thumb, the opening of the pseudospectral lines designates spectral instability, which is justified here due to the non-normality of the respective operator. Take for example the second contour line from the top on the left panel of Fig.~\ref{spectral instability}. This roughly corresponds to a perturbation in $L$ of order $\sim 10^{-25}$. Nevertheless, from the 10th mode and above, the eigenvalues have the freedom to migrate at any point from the specific contour and above boundlessly and disproportionately with respect to the scale of the perturbation. The contours discussed are therefore the bound of eigenvalue migration below which the remaining eigenvalues remain stable under the particular perturbation of order $\epsilon$. Even so, a larger norm perturbation can further destabilize more overtones, and even the fundamental mode as the pseudospectrum entails. Notice that we only show a part of the complex plane; the contour lines extend far and wide and behave smoothly as long as the number of interpolation grid points used is large enough.

An important aspect of the results shown in the left panel of Fig.~\ref{spectral instability} is that the assumed perturbation is not only constrained to the part of the operator that contains the effective potential, but rather affects $L$ in its entirety. This result, although non-trivial and interesting for other systems, is not straightforwardly connected to perturbations of the effective potential of a BH due to environmental effects, quantum corrections, and other effects. To understand if the pseudospectrum is a meaningful tool for BH physics, the authors in Ref. \cite{Jaramillo2020} perturbed only the P\"oschl-Teller potential part of $L_2$, by adding two different kinds of energy norm perturbations: (i) random perturbations to the potential, designated as $||\delta V_r||_E$ and (ii) deterministic perturbations designated as $||\delta V_d||_E\sim \cos(2\pi kx)$, where $k$ is the frequency of the perturbation. Both types of perturbations are normalized such that their absolute value in the energy norm is a dimensionless number. By merely choosing a random perturbation of order $10^{-16}$, as shown in the right panel of Fig.~\ref{spectral instability}, we observe that from the 7th mode and on, a QNM migration, that practically lies on top of the respective pseudospectral contour lines, occurs. Increasing the order of the perturbation destabilizes even lower overtones. Since random additions of numbers on the effective potential do not necessarily have a physical meaning, the authors in~\cite{Jaramillo2020} also included deterministic perturbations by adding a sinusoid with some scale and frequency. This is a more physical scenario of a perturbing agent such as a quantum effect. The results in this case are also fascinating: by fixing the order of the perturbation to $10^{-8}$ and choosing a small fluctuation frequency $k$, there is no observed migration, and the system may falsely seem to exhibit spectral stability. Nevertheless, the increment of $k$ has the same effect with random perturbations, where larger frequencies induce destabilization of lower overtones. Significantly, no matter the order of random and the order/frequency combinations of deterministic perturbations, the fundamental mode remains stable to these fluctuations, even though the pseudospectrum entails that even the fundamental mode can migrate for appropriate $\epsilon$. This observation practically designates the difference between perturbing only the potential of the system or perturbing its whole operator. Nonetheless, the fact that a tiny addition to the effective potential can lead to the destabilization of overtones is fundamental for performing validity tests of GR and spectral instabilities, if present in BHs, need to be taken seriously for the BH spectroscopy program to be successful.

\subsection{Schwarzschild black holes}

Schwarzschild BHs are exact static and spherically-symmetric solutions of the vacuum Einstein field equations. The line element describing them has the form of Eq.~\eqref{line_element} with
\begin{equation}
	f(r)=1-\frac{2M}{r},
\end{equation}
where $M$ is gravitational mass of the object. They possess an event horizon at $r=2M$, which satisfies $f(2M)=0$. 

Scalar, electromagnetic and axial gravitational perturbations are described by Eq.~\eqref{wave_equation_time}. The axial gravitational effective potential of a perturbed Schwarzschild BH, known as the Regge-Wheeler potential is isospectral (same QNM spectrum) with the polar (Zerilli) sector, and has the form
\begin{equation}\label{axial potential}
	V=\left(1-\frac{2M}{r}\right)\left(\frac{\ell(\ell+1)}{r^2}-\frac{6M}{r^3}\right).
\end{equation}
The potential in Eq.~\eqref{axial potential} decays exponentially at the event horizon but in contrast to the P\"oschl-Teller function, it decays polynomially at infinity, thus it includes a branch cut in the complex plane of the Green's function. In this case, the counterpart of the power-law cutoff in the hyperboloidal formulation used in~\cite{Jaramillo2020} gives rise to a continuous part in the spectrum, besides the typical convergent QNMs, that is not converging as $N$ increases. These continuous eigenvalues are not QNMs and can be unabiguously identified, though their presence must be considered when computing the pseudospectra. Thus, the benchmark case of the P\"oschl-Teller potential will prove crucial for realistic scenarios.

\subsubsection{Pseudospectrum of Schwarzschild black holes and perturbed quasinormal modes}

\begin{figure}[hb]\centering
	\includegraphics[scale=0.42]{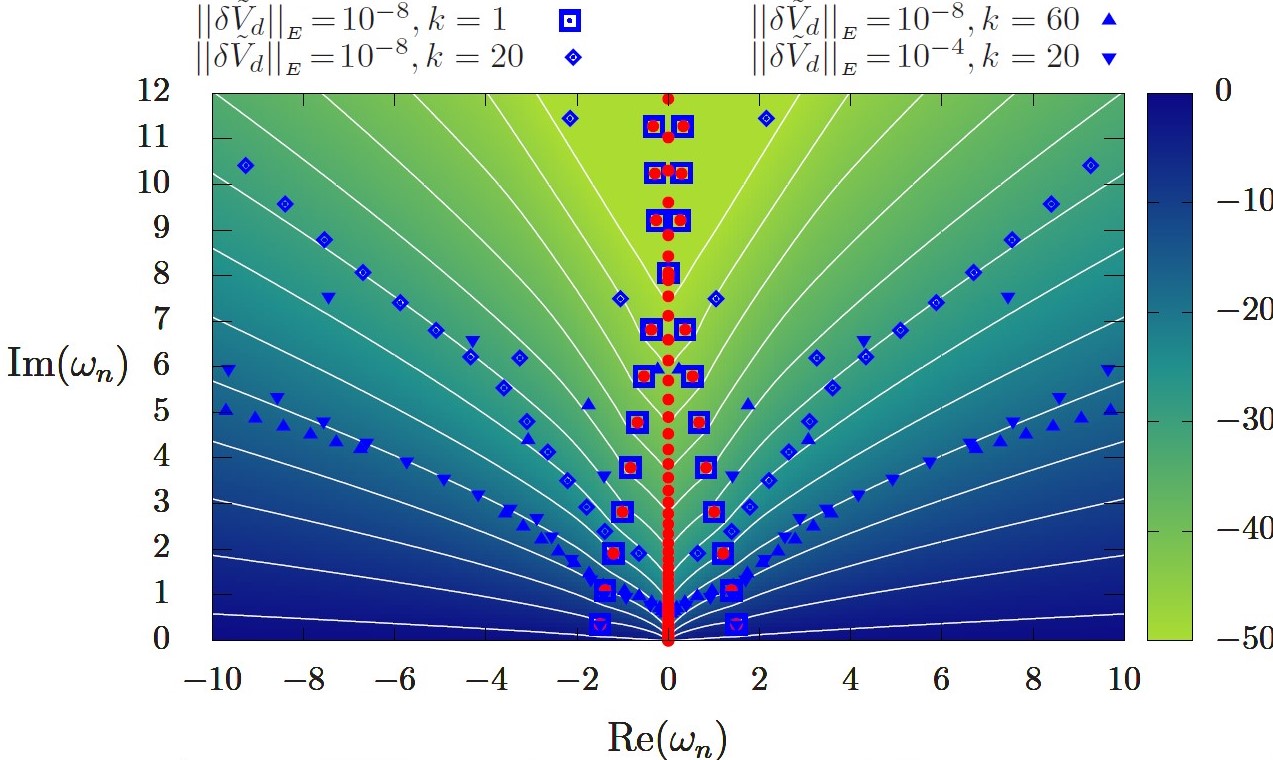}
	\caption{Pseudospectrum of Schwarzschild BH QNMs $N=200$ grid points. The white lines correspond to different $\epsilon$ pseudospectral contour lines which become more negative close to the eigenvalues (red dots). The colorbar on the right in is logarithmic scale meaning that a value of, e.g. $-30$ corresponds to $\log[\sigma^{-30}(L)]$ or simply a perturbation in $L$ with energy norm $10^{-30}$. On top of the pseudospectrum, perturbed QNMs resulting from deterministic $\delta V_d$ perturbations to the BH potential are plotted with blue markers that vary according to the oscillation frequency $k$ of the perturbation. They are normalized such that in the energy norm they acquire the specified normalized value shown above the figure. The figure has been adapted from~\cite{Jaramillo2020}.}
	\label{Schwarz_pseudospectrum}
\end{figure}

Again, by applying the methods outlined in section~\ref{pseudo_calculation}, we can bring the wave equation, with $V$ as in Eq.~\eqref{axial potential}, into a matrix evolution form, that allows us to find the QNMs and pseudospectra of Schwarzschild BH QNMs (see~\cite{Jaramillo2020} for a detailed description of the steps and functions involved). Besides the pseudospectral analysis, the steps of adding sinusoid perturbations to the BH potential is repeated to demonstrate that the pseudospectra contour lines correspond to the destabilization boundaries of perturbed QNMs above which they can migrate in the complex plane disproportionally. Figure~\ref{Schwarz_pseudospectrum} conforms with the above discussion, i.e. the pseudospectral contour lines for different level sets $\epsilon$ extend to the edges of the studied range of the complex plane in a logarithmic manner, as in the P\"oschl-Teller case. A significant difference here is the appearance of eigenvalues along the imaginary axis. These modes are non-convergent as $N$ is increased, they are continuous and they are associated to the branch-cut in the complex plane of solutions of the Green's function~\cite{Leaver86c} that leads to the inverse-polynomial decay of perturbations at infinity~\cite{Price:1971fb,Price:1972pw,Gundlach:1993tp,Gundlach:1993tn}. Even though they do not belong in the QNM spectrum of Schwarzschild BHs, they significantly affect the pseudospectrum close to the imaginary axis by bending it further downwards. This is in contrast to the P\"oschl-Teller case where the potential decays exponentially on both sides and the branch-cut is absent, thus a smoother pseudospectrum is obtained even at the imaginary axis. These non-convergent modes pose an open issue regarding their existence in asymptotically flat BHs and may be connected to the hyperboloidal approach, which explicitly includes null infinity into the grid. Nevertheless, they do not taint the underlying migration of perturbed QNMs since the opening of the contour lines is a clear-cut imprint of spectral instability.

It is important to note that besides the random perturbations that were utilized in P\"oschl-Teller potentials, deterministic perturbations for the Schwarzschild case are more physically relevant for two reasons. Firstly, the results are qualitatively the same as those arising from random perturbations. Secondly, deterministic sinusoidal perturbations to the effective potential may have an astrophysical relevance to explore the effects of perturbations (assessment of ``long range/low frequency'' versus ``small scale/high frequency'' perturbations), as well as those arising from quantum gravity phenomenology (``small scale/high frequency'' effective fluctuations). 

In Fig.~\ref{Schwarz_pseudospectrum}, besides the pseudospectrum, four cases of different combinations of perturbation norm and frequency are over-plotted, shown with different blue markers. The perturbed spectra confirm once more that perturbations to the potential almost coincide with the pseudospectra contour lines following the behavior found in~\cite{Nollert:1996rf} (shown in Fig.~\ref{Nollert_branch}). Yet, there is a crucial difference. The fundamental mode is never destabilized, no matter the scale or frequency of the perturbation, in contrast to what the pseudospectrum suggests for sufficiently large $\epsilon$. The same holds for random perturbations. On the other hand, overtones are prone to migrate on the branches of maximal migration as the perturbation frequency increases. This provides a definitive connection of perturbed QNMs, resulting by adding perturbations to the Schwarzschild potential, and pseudospectra branches, which led the authors in~\cite{Jaramillo2020} to conclude that perturbed QNMs track pseudospectra contour lines, thus ``christening'' them ``Nollert-Price'' branches.

\subsection{Reissner-Nordstr\"om black holes}

RN BHs are exact solutions of the Einstein-Maxwell equations that include, besides the BH mass $M$, an electric charge $Q$. In this case, the lapse function takes the form
\begin{equation}
	f(r)=1-\frac{2M}{r}+\frac{Q^2}{r^2}.
\end{equation}
RN BHs possess two horizons, an event horizon at $r=r_+$ and an inner (Cauchy) horizon $r_-$, beyond which the deterministic nature of the field equations is lost~\cite{Hawking:1973uf,Cardoso:2017soq}. Their exact forms are
\begin{equation}
	r_\pm=M\pm\sqrt{M^2-Q^2}.
\end{equation}
RN BHs allow us to perform further tests of geometrical aspects of pseudospectra universality, which could not be performed with Schwarzschild BHs. Due to the existence of two horizons, there exist two choices of spacetime slicings for the hyperboloidal approach~\cite{PanossoMacedo:2018hab}, namely the \emph{areal radius fixing gauge}, where we fix $r=r_+/x$\footnote{We remind that $x\in[0,1]$ is the appropriately compactified radial coordinate through a choice of $g(x)$.} and the \emph{Cauchy horizon fixing gauge}, where we fix $x_-\rightarrow\infty$, that corresponds to the location of the Cauchy horizon at the compactified coordinate. The hyperboloidal approach can be applied in both slicings, and tests have shown that different gauges yield the same results, supporting the geometrical nature of the pseudospectrum~\cite{Destounis:2021lum}. Furthermore, RN BHs possess an extremal limit when $Q=M$, in which $r_+=r_-$. Approaching this limit gives rise to a new family of near-extremal, purely imaginary and long-lived modes as $Q\rightarrow M$~\cite{Kim:2012mh,Zimmerman:2014aha,Richartz:2015saa,Cardoso:2017soq,Cardoso:2018nvb,Destounis:2018qnb,Liu:2021aqh}. At exact extremality, RN BHs become marginally stable~\cite{Aretakis:2011ha,Aretakis:2011hc} and allow for the formation of local horizon hair~\cite{Angelopoulos:2018yvt}. Finally, the gravitational-led QNMs with angular number $\ell$ are isospectral with electromagnetic-led QNMs with angular number $\ell-1$ when the BH is extremal~\cite{Onozawa:1996ba,Okamura:1997ic,Kallosh:1997ug,Berti:2004md}.

Here, we focus on scalar and polar gravitoelectric perturbations, described by the potential~\cite{Moncrief74a,Moncrief74b,Moncrief75,Destounis:2021lum}
\begin{equation}
	V= \dfrac{f(r)}{r^2}\left[ \ell(\ell+1) + \dfrac{r_+}{r}\left(\mu - \kappa^2 \nu \dfrac{r_+}{r}\right)  \right],\label{eq:potential_RN}
\end{equation}
with
\begin{align}\label{potential factors}
	\nu &= 3 n_p - 1, \quad \kappa=Q/r_+=\frac{Q/M}{1+\sqrt{1-(Q/M)^2}} \quad \mu = n_p(1+\kappa^2) - (1-n_p)\mathfrak{m}_\pm, \nonumber\\
	\mathfrak{m}_\pm &= \dfrac{1+\kappa^2}{4} \Bigg[1 \pm 3 \sqrt{1 + \dfrac{4\kappa^2 A}{(1+\kappa^2)^2}}\Bigg], \quad A = \dfrac{4}{9}(\ell+2)(\ell-1).\nonumber
\end{align}
In the above parameterization of equations, scalar perturbations are recovered for $n_p=1$, whereas electromagnetic-led $(\mathfrak{m}_-)$ and gravitational-led $(\mathfrak{m}_+)$ perturbations (which reduce to electromagnetic and gravitational perturbations of Schwarzschild BHs in the uncharged limit) are recovered for $n_p=-1$. 

\subsubsection{Pseudospectrum of Reissner-Nordstr\"om black holes}

By using the areal radius fixing gauge, we have calculated the spectra and pseudospectra of RN BHs, for various type of perturbations with a constant grid resolution of $N=200$ Chebyshev points. Due to the asymptotic flatness of spacetime, similar non-convergent eigenvalues appears in the imaginary axis which from now on we designate with blue dots for clarity. For more details regarding the functions and gauges involved for the hyperboloidal approach, as well as the numerical calculation of the pseudospectrum in the energy norm, we refer the reader to~\cite{Destounis:2021lum}.

\begin{figure}\centering
	\includegraphics[scale=0.32]{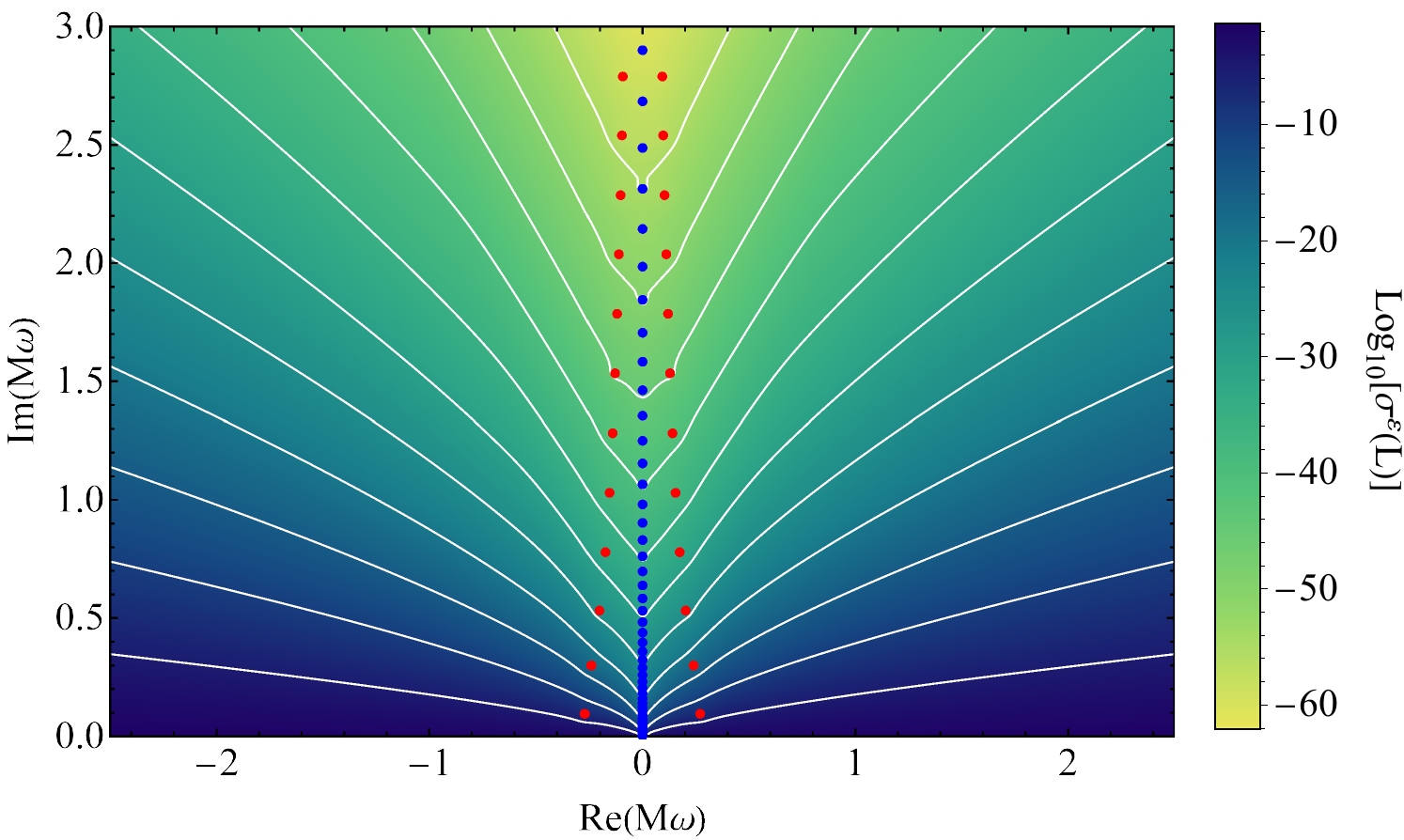}\hspace{0.5cm}
	\includegraphics[scale=0.32]{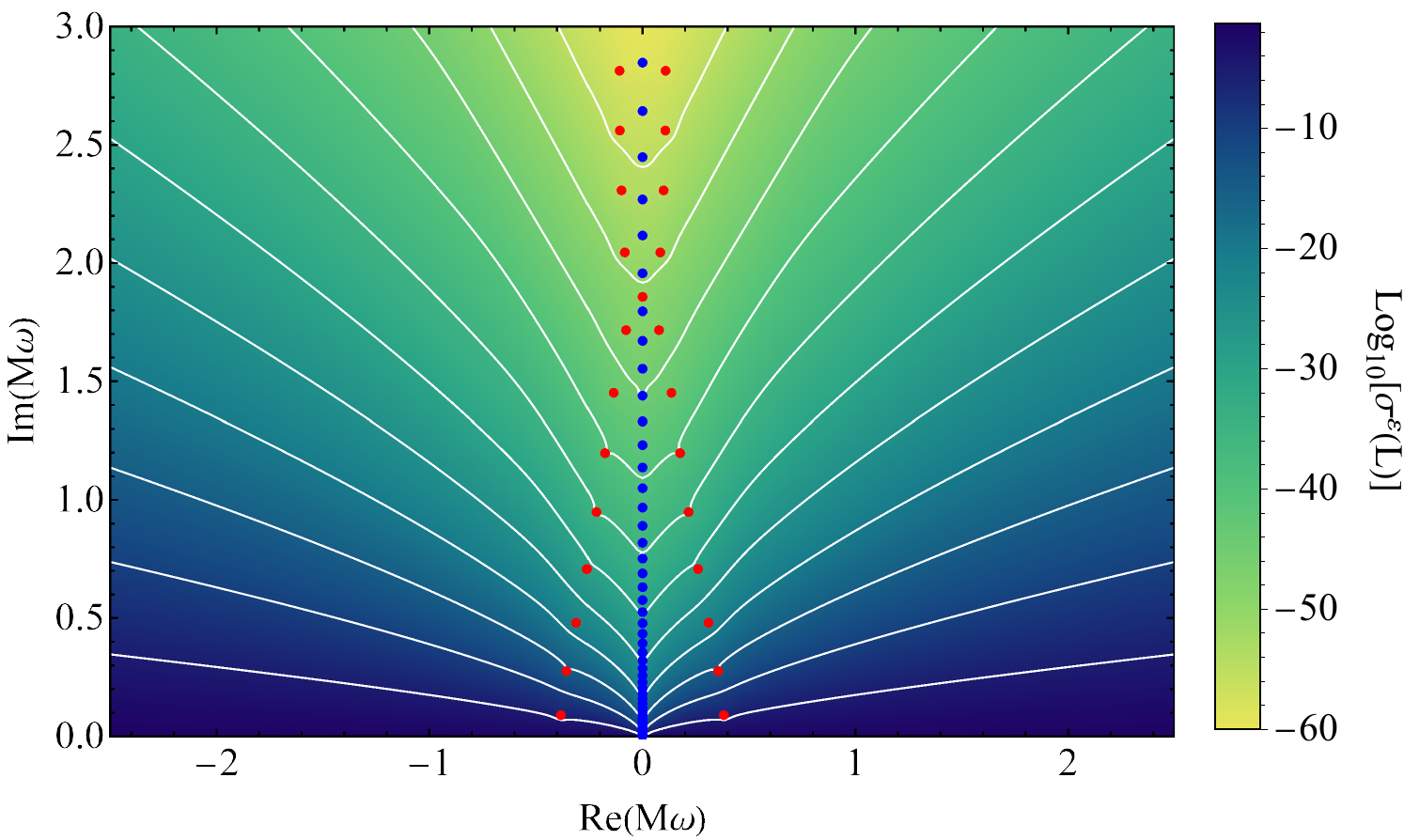}
	\caption{Pseudospectra of a BH with charge $Q/M=0.5$. Left: $\ell=1$ electromagnetic-led QNMs and $\epsilon$-pseudospectra boundaries. Right: Same for $\ell=2$ gravitational-led QNMs. The contour levels $\log_{10}(\epsilon)$ range from $-55$ (top level) to $-5$ (bottom level) with steps of $5$. The blue dots designate branch-cut (non-convergent) modes while the red dots designate QNMs. The figure has been taken from~\cite{Destounis:2021lum}.}
	\label{grel}
\end{figure}

In Fig.~\ref{grel} we show the electromagnetic-led and gravitational-led spectra and pseudospectra of a RN BH with charge $Q/M=0.5$. The pseudospectrum is qualitatively similar to those obtained for Schwarzschild BHs (see Fig.~\ref{Schwarz_pseudospectrum}), with pseudospectral contour lines designating a spectral instability of QNMs and a logarithmically-extending pattern away from QNMs. In fact, according to~\cite{Jaramillo:2021tmt}, the QNM-free regions are asymptotically bounded from below by logarithmic curves of the form
\begin{equation}
	\text{Im}(\omega) \sim C_1+C_2\log [\text{Re}(\omega)+C_3].
\end{equation}
Remarkably, appropriate choices of the constants $C_1,\,C_2$ and $C_3$ can lead to a fit that holds also at intermediate values of $\text{Re}(\omega)$ and even close to the QNMs; a region one can hardly consider as asymptotic.

\subsubsection{The extremal limit}

By examining the spectra and pseudospectra of extremal RN for scalar perturbations we find that their overall behavior is the same as the one found for gravitoelectric pseudospectra.

\begin{figure}[b!]\centering
	\includegraphics[scale=0.34]{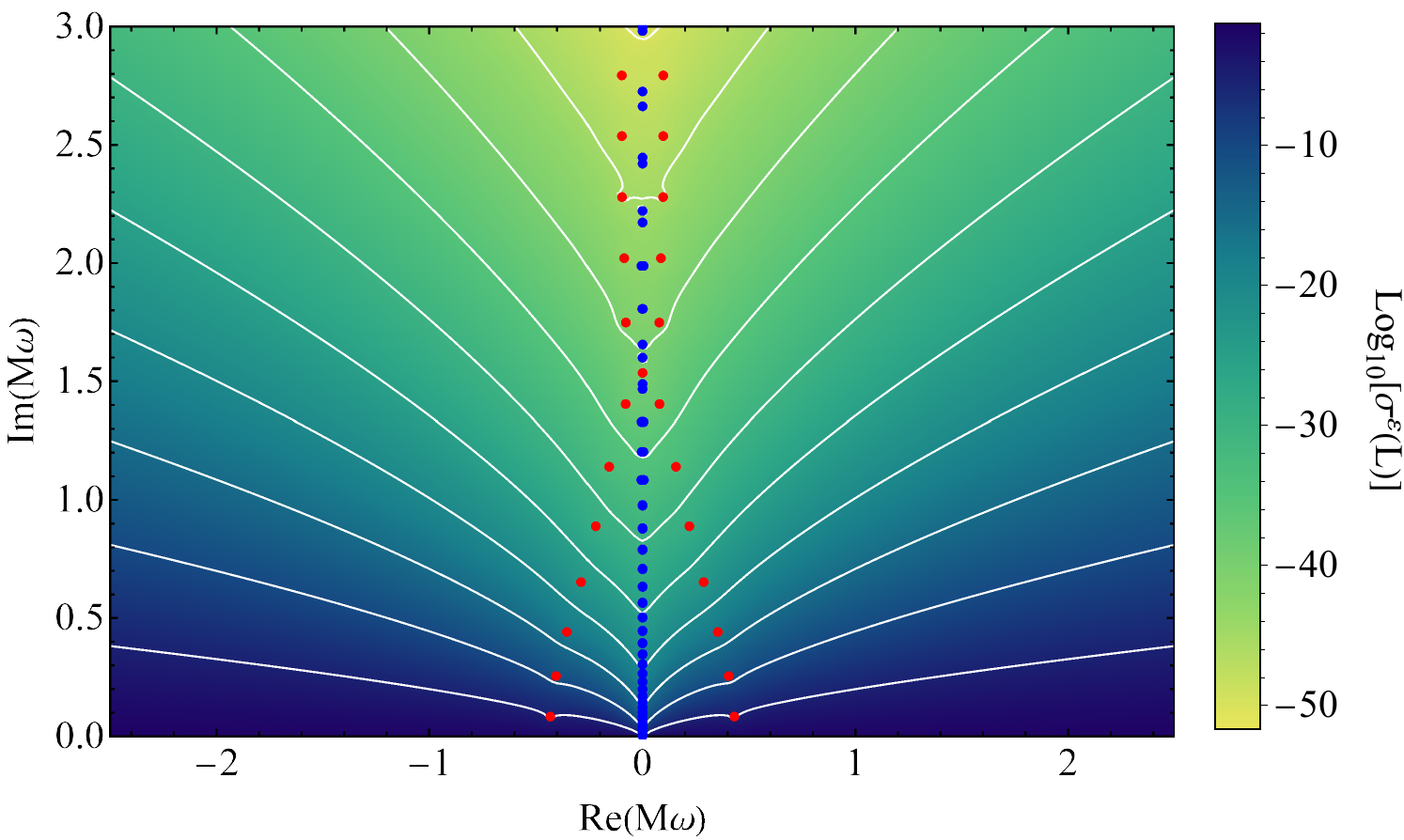}\hspace{0.5cm}
	\includegraphics[scale=0.39]{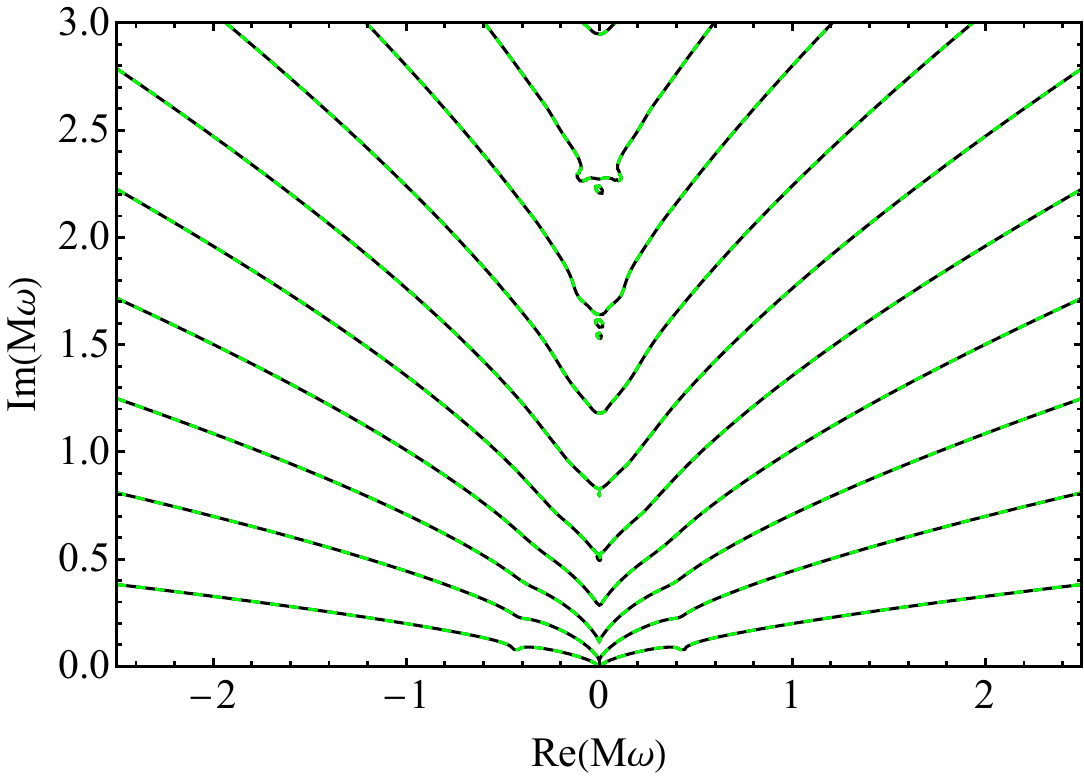}
	\caption{Left: $\ell=2$ gravitational-led QNMs (red dots) and $\epsilon$-pseudospectra boundaries of a RN BH with $Q/M=1$. Right: superimposed $\ell=1$ electromagnetic-led (black lines) and $\ell=2$ gravitational-led (green dashed lines) $\epsilon$-pseudospectral contours of a RN BH with $Q/M=1$. In both cases, the contour levels $\log_{10}(\epsilon)$ range from $-50$ (top level) to $-5$ (bottom level) in steps of $5$. The blue dots designate branch-cut (non-convergent) modes. The figure has been taken from~\cite{Destounis:2021lum}.}
	\label{grelext}
\end{figure}

The pseudospectrum in extremality possesses logarithmic asymptotics and exhibits a zero-frequency mode ($M\omega=0$), associated with the Aretakis instability~\cite{Aretakis:2011ha,Aretakis:2011hc}. In general, modes that reside close to the real axis are significant to transient instabilities that are typically resolved with pseudospectra, as discussed in Section~\ref{hydro}. Nevertheless, the existence of branch-cut modes that infect the spectrum with noise, and subsequently the pseudospectrum at the imaginary axis, forbids us from drawing any solid conclusions regarding transient phenomena in extremal spacetimes, since the crossing of the contour lines to the side of the complex plane where unstable modes reside may be tainted by the branch-cut modes.

For gravitoelectric QNMs, the $\ell-1$ electromagnetic-led and $\ell$ gravitational-led modes of extremal RN geometries are isospectral \cite{Berti:2004md,Onozawa:1996ba}. Figure~\ref{grelext} shows that not only the spectra coincide, but also the pseudospectra for such perturbations at extremality. Thus, the isospectrality taking place here has a stronger nature and occurs because not only the poles, but the Green's function as a whole coincide in the two cases.

\begin{figure}[t]\centering
	\includegraphics[scale=0.43]{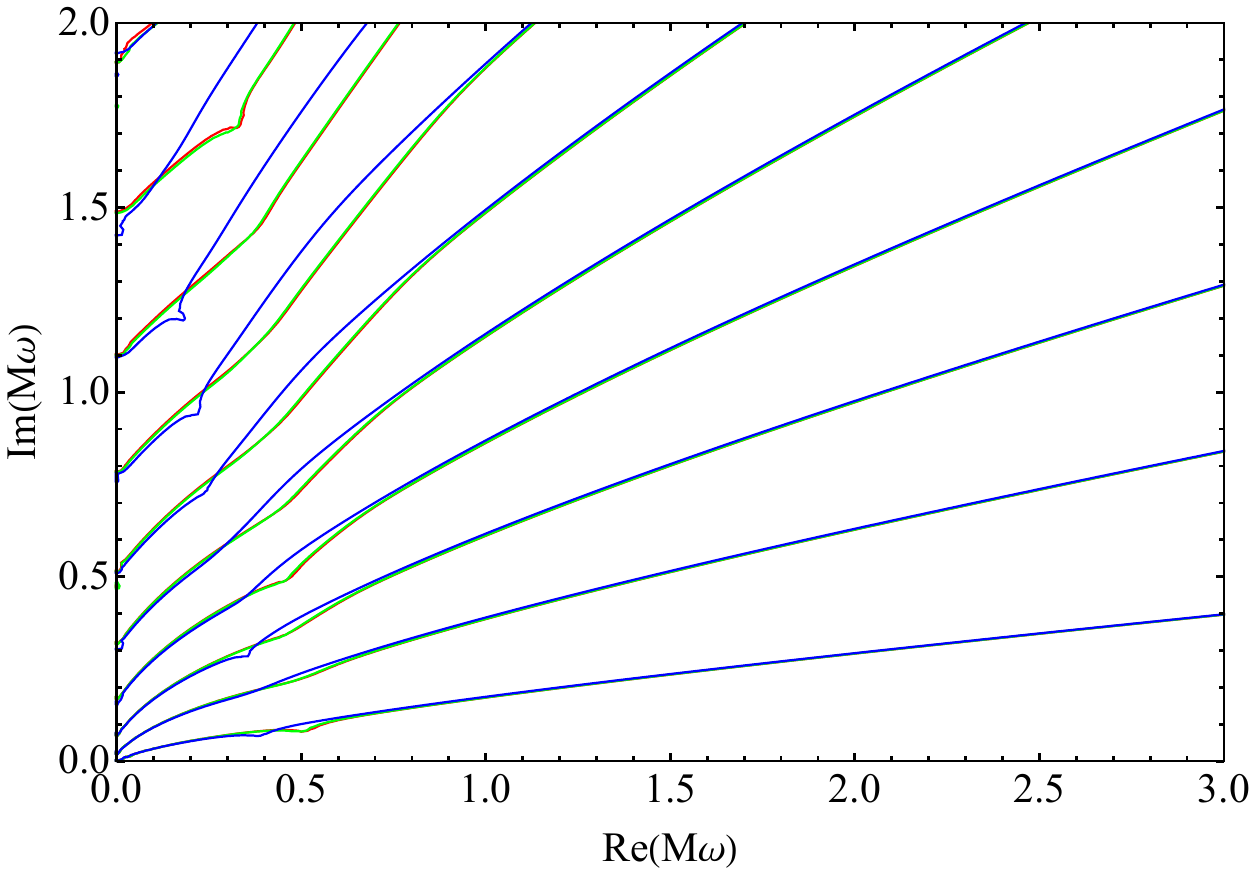}
	\caption{Superimposed scalar (red), electromagnetic-led (green) and gravitational-led (blue) pseudospectral levels for a RN BH with $Q/M=0.5$. All perturbations share the same angular index $\ell=2$. The contour levels $\log_{10}(\epsilon)$ range from $-50$ (top level) to $-5$ (bottom level) with steps of $5$. The figure has been taken from~\cite{Destounis:2021lum}.}
	\label{asympt}
\end{figure}

\subsubsection{Asymptotic universality of pseudospectra}

We finally discuss the asymptotic behavior of pseudospectra for different kinds of fluctuations. Figure~\ref{asympt} shows the asymptotic structure of three spectral problems, namely the $\ell=2$ scalar, electromagnetic-led and gravitational-led QNMs of a subextremal RN BH. It is evident that the lower pseudospectra contours of the three distinct effective potentials coincide already at small frequencies and almost from the respective QNM for each level set, while higher contours begin to coincide at frequencies with larger real parts. This poses as a numerical evidence of the asymptotic universality of pseudospectra shared by a whole class of BH potentials, with similar logarithmic patterns.

\section{Pseudospectrum of horizonless exotic compact objects}

Our current understanding of stellar evolution suggests that astrophysical compact objects can experience the process of gravitational collapse when the interaction of stellar matter cannot counterbalance the star's self-gravity. The most natural outcome is a collapse to a BH. Nevertheless, the singularity that appears through this classical process may hint at an incomplete model of gravitational collapse. In order to avoid singularities, we can consider either localized corrections around the singularity, leaving the exterior of the BH effectively classical, or slightly larger scale corrections that are still localized but can extend slightly outside the event horizon, thus removing the event horizon. Currently, a plethora of horizonless ECOs has been constructed through the latter aforementioned technique, such as classical boson stars~\cite{Liebling:2012fv,Vaglio:2022flq,Destounis:2023khj}, gravastars~\cite{Mazur:2001fv}, or semi-classical ECOs with exotic near-horizon structure~\cite{Carballo-Rubio:2017tlh,Arrechea:2021pvg} such as firewalls~\cite{Kaplan:2018dqx} and fuzzballs~\cite{Mathur:2005zp,Mathur:2008nj,Ikeda:2021uvc}.

ECOs are interesting not only due to their composition, but also their dynamical behavior under perturbations. Their QNMs can differ dramatically from those of BHs, due to the different boundary condition that has to be imposed at their surface. In particular, QNMs of BHs require purely ingoing waves at the event horizon while for horizonless ECOs that possess a centrifugal barrier, slightly outside the would-be event horizon, the boundary condition may become (at least partially) reflective. This in turn, can lead to fundamental modes that are extremely long-lived, when the ECO is ultracompact, which are clearly distinguishable from those of BHs~\cite{Barausse:2014tra,Cardoso:2016rao,Cardoso:2016oxy,Cardoso:2019rvt}. These long-lived QNMs, which are obtained with BH perturbation theory at the linear level, can source instabilities in ECOs when non-linear effects are taken into account~\cite{Cardoso:2014sna,Keir:2014oka,Cunha:2022gde}. Thus, constructing stable ultracompact horizonless objects that can mimick BHs seem to be a non-trivial task. 

The motivation for studying the pseudospectrum of spherically-symmetric horizonless ECOs stems from the fact that they both exhibit arbitrarily long-lived QNMs (that reside close to the real axis of the complex plane) and, at the non-linear level, they become unstable due to various instability mechanisms \cite{Zhong:2022jke,Cunha:2022gde}. Such a problem is in total analogy to the results discussed in Section~\ref{hydro} regarding hydrodynamic flows, where a linear analysis fails to capture the phenomena found in experiments, and the pseudospectrum resolved the inconsistency between linear analyses and full non-linear experiments.

\subsection{Spherically-symmetric horizonless compact objects: a simplistic approach}

Even though there are plenty of ECO configurations that can be constructed for the purpose of studying their QNMs and pseudospectra, such a task is highly demanding. As in neutron stars, a complete ECO model requires knowledge of the object's internal structure and how its matter content interacts with external fluctuations. In~\cite{Boyanov:2022ark}, we have instead taken a simplified approach to construct a spherically-symmetric ECO effectively by hand that reproduces the whole phenomenology of realistically constructed ECOs. Our analysis assumes a Schwarzschild BH spacetime on the domain $r\in[r_0,\infty)$, where $r=r_0=(1+\mathcal{E})r_+$ is the ECO surface and $\mathcal{E}\ll1$ for ultracompact objects. At $r=r_0$ we impose a purely reflective boundary condition, i.e. for a perturbation field $\Psi$ we assume the Dirichlet condition $\Psi(t,r_0)=0$. Even though our assumption of pure reflection of waves at the ECO surface is very simplistic, there are other choices where one can introduce partial reflection and absorption~\cite{Maggio:2020jml,Vellucci2022,Zhong:2022jke}. In the interest of studying the spectrum and pseudospectrum on a static background that does not absorb energy, we will assume a purely reflective ECO surface, though the results that follow can be extrapolated to other choices of partially reflective boundary conditions. In the frequency domain, where our results will be interpreted, the boundary condition at infinity remains the same as that of an asymptotically-flat BH, i.e. purely outgoing waves, while at the surface of the ECO we impose a Dirichlet boundary condition $\Psi(r_0)=0$. 

The majority of the hyperboloidal approach and the energy norm is used exactly as described in Section~\ref{pseudo_calculation}. Specifically, at null infinity, the outgoing boundary condition is imposed automatically. At the ECO surface though, the purely reflective boundary condition is achieved by removing the row and column of the discretized operator $L$ that corresponds to $r=r_0$, or equivalently $x=x_0$ at the compactified coordinate system utilized for numerical calculations. For a detailed analysis of the equations involved to properly set up the wave equation of perturbations of ECOs in hyperboloidal coordinates, we refer the reader to~\cite{Jackiewicz:2002,Boyanov:2022ark}.

\subsection{Pseudospectrum of horizonless objects with a P\"oschl-Teller potential}\label{PT-ECO}

\begin{figure}[b!]
	\centering
	\includegraphics[scale=0.825]{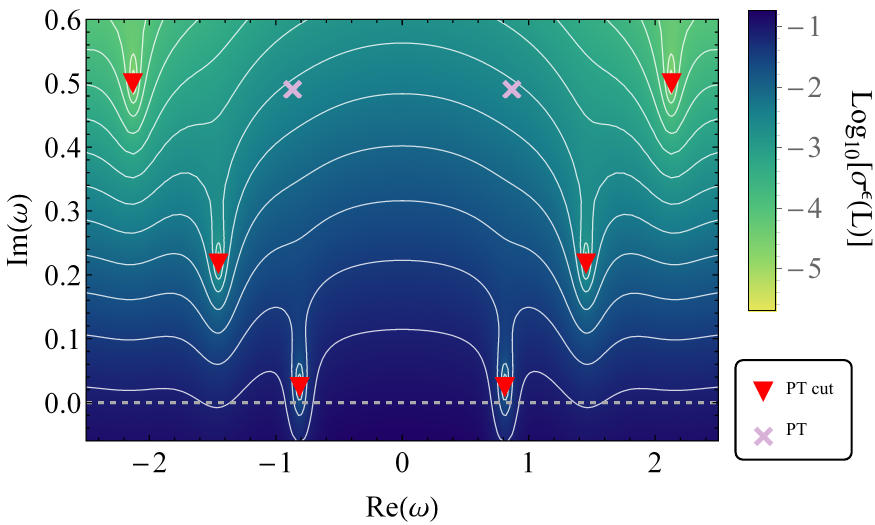}
	\caption{QNMs (red triangles) and pseudospectrum (white curves) of the $L$ operator for the P\"oschl-Teller potential with a Dirichlet boundary condition at $\mathcal{E}=10^{-3}$ (PT cut) and $N=200$ spatial discretization points. Contour lines correspond to values between $-4.7$ (uppermost contour) and $-1.1$ (lowest contour) with steps of $0.3$ in the log scale. The horizontal dashed line indicates the position where $\text{Im}(\omega)=0$. For comparison, we also include the QNMs of the P\"oschl-Teller with outgoing boundaries on the right and ingoing on the left of the domain (PT, pink crosses). The figure has been adapted from~\cite{Boyanov:2022ark}.}
	\label{PTps}
\end{figure} 

Following the steps of~\cite{Jaramillo2020}, we first analyze a simple case where the ECO potential is replaced with the P\"oschl-Teller function~\eqref{PT_potential}, where at the right of the potential, that corresponds to infinity, the boundary condition is kept outgoing and at the left of the potential we use a Dirichlet boundary at a finite distance $x=x_0$ where the hypothetical ECO surface resides. In Fig.~\ref{PTps}, we show the spectra and pseudospectra resulting from the aforementioned system. It is apparent that the QNMs of the P\"oschl-Teller ECO are much more long-lived than those of the P\"oschl-Teller potential itself, which is an illustration of stable trapping that occurs in ECOs. Moreover, ECO modes suffer from spectral instability, like any other dissipative system discussed. Most importantly, the pseudospectrum suggests that the fundamental QNM and first overtone are able to cross the real axis, into the unstable part of the complex plane, for $\epsilon\simeq 0.02$. This is suggestive of a perturbation-triggered instability, where when one adds a random or deterministic perturbation into the effective potential, the long-lived modes may migrate to the lower half of the complex plane where unstable QNMs reside. Nevertheless, by employing the aforementioned strategy, we reached the same conclusions as those for BHs: the fundamental mode is spectrally stable as long as the asymptotic structure of the potential remains intact.

Despite that, the protrusion of the pseudospectral contour lines into the unstable part of the complex plane indicates a tendency with which pseudoresonances can be triggered easily, such as those found in hydrodynamics~\cite{Trefethen:1993}, if the evolution operator had a time-dependent external source, whose Fourier transform had a real frequency close to the pseudospectrum protrusion regime~\cite{Jaramillo:2022kuv}.

\subsection{Pseudospectrum of horizonless compact objects}

The analysis in Section~\ref{PT-ECO} has been repeated for the scalar and axial gravitational perturbation potentials of a Schwarzschild BH, where instead of ingoing waves at the event horizon, we impose a Dirichlet boundary at the surface of the ECO $r=r_0$. In this case, $\mathcal{E}$ controls the compactness of the ECO. The smaller it gets, the closer the ECO surface gets to the would-be event horizon, and the object becomes more compact. This allows for perturbations to be trapped between the photon sphere and the ECO surface, leading to long-lived modes in the frequency domain and echoes in the ringdown signal~\cite{Cardoso:2016rao,Cardoso:2016ryw,Cardoso:2017cqb,Abedi:2020ujo}. The systematics of calculating the QNMs and pseudospectra are detailed in~\cite{Boyanov:2022ark} and in the previous sections.

\begin{figure}
	\centering
	\includegraphics[scale=0.55]{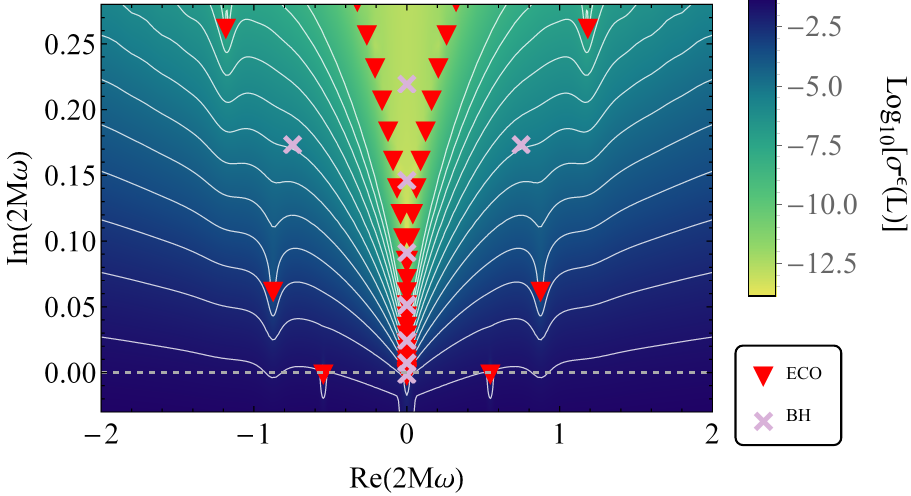}\hspace{0.25cm}
	\includegraphics[scale=0.55]{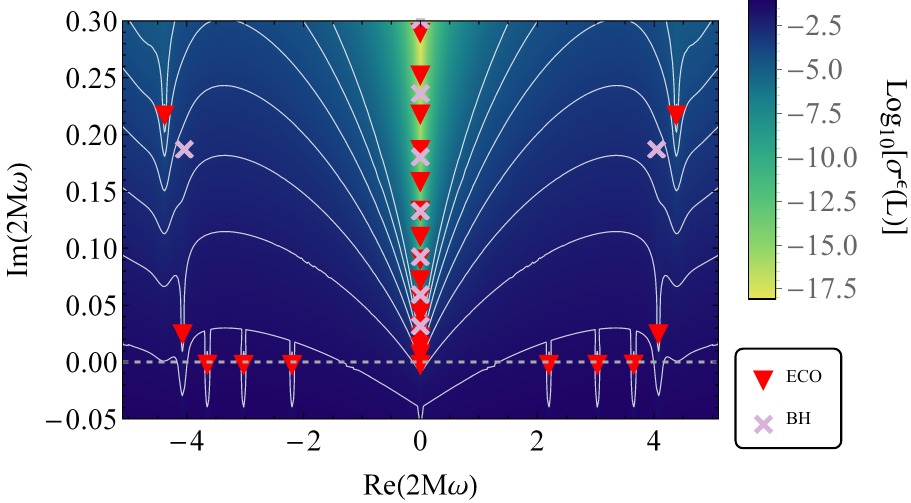}
	\caption{Left: ECO axial gravitational $\ell=2$ QNMs (ECO, red triangles) with compactness $\mathcal{E}=10^{-3}$. The contour lines range from $-6.5$ (uppermost contour) to $-1.7$ (lowest contour) with steps of $0.6$ in the log scale. Right: ECO scalar $\ell=10$ QNMs (ECO, red triangles) with compactness $\mathcal{E}=10^{-2}$. The contour lines range from $-6.5$ (uppermost contour) to $-1.7$ (lowest contour) with steps of $0.6$ in the log scale. In both figures, we also include the respective field QNMs of a Schwarzschild BH (BH, pink crosses). The figure is adapted from~\cite{Boyanov:2022ark}.}
	\label{ECO_PS_nopert}
\end{figure}

Figure~\ref{ECO_PS_nopert} demonstrates, as expected, similar results as those found in the P\"oschl-Teller case discussed above. Scalar and axial gravitational QNMs of ECOs with varying compactness $\mathcal{E}$ suffer from spectral instabilities, as their pseudospectra suggest, and the long-lived QNMs close to the real axis protrude to the unstable side of the complex plane. Therefore, with a similar mechanism as in the P\"oschl-Teller potential, a time-dependent source with a real frequency in the vicinity of the protrusion of the contour lines could trigger a pseudoresonance that can, in turn, lead to an instability, in a similar manner as it is found for ultracompact objects when higher than linear-order perturbations are taken into consideration~\cite{Cardoso:2014sna,Keir:2014oka,Cunha:2022gde,Zhong:2022jke}.

\begin{figure}[t]
	\centering
	\includegraphics[scale=0.555]{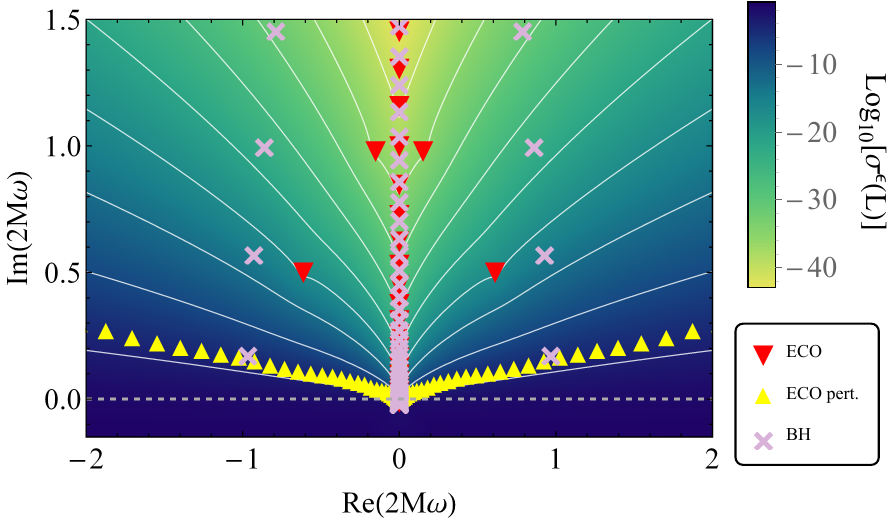}\hspace{0.25cm}
	\includegraphics[scale=0.555]{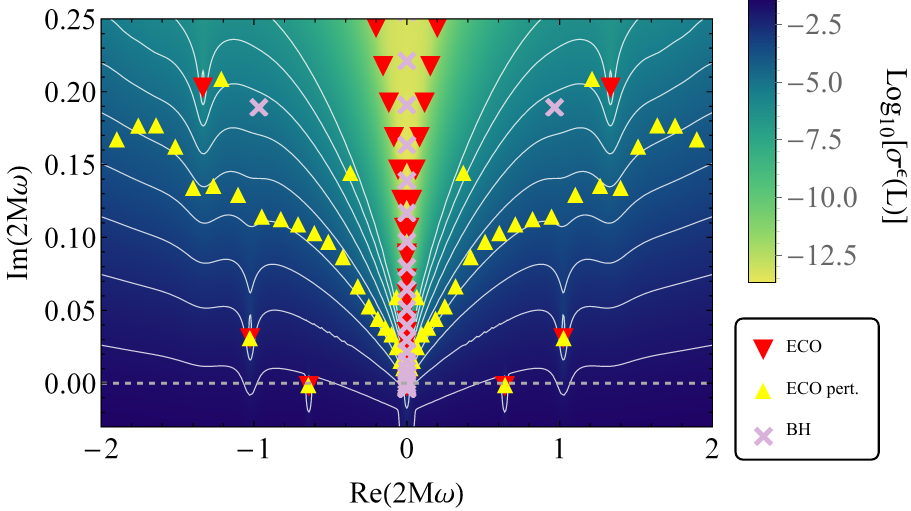}
	\caption{Left: ECO scalar $\ell=2$ QNMs (ECO, red triangles) with compactness $\mathcal{E}=1$, and the perturbed QNMs (ECO pert., yellow triangles) resulting by adding a random perturbation to the effective potential of energy norm $10^{-2}$. The contour lines range from $-44$ (uppermost contour) to $-3.5$ (lowest contour) with steps of $4.5$ in the log scale. Right: Same as left with compactness $\mathcal{E}=10^{-3}$, and a random perturbation to the effective potential of energy norm $10^{-1}$. The contour lines range from $-6.5$ (uppermost contour) to $-1.7$ (lowest contour) with steps of $0.6$ in the log scale. On both figures, we also include the scalar $\ell=2$ QNMs of a Schwarzschild BH for comparison (BH, pink crosses). The figure is adapted from~\cite{Boyanov:2022ark}.}
	\label{ECOps}
\end{figure}

Nonetheless, when adding random perturbations to the effective potentials of scalar and axial gravitational perturbations of the ECO we find interesting results. For less compact objects, such as $\mathcal{E}=1$ (see Fig.~\ref{ECOps}), the ECO QNMs have an imaginary part that is larger than the fundamental BH mode with the same mass. As the compactness increases (decreasing $\mathcal{E}$), the imaginary parts of the ECO QNMs become significantly smaller than those of a BH. We find the same results for both types of field perturbations. Once more, stable trapping, and the formation of quasibound states~\cite{Dolan:2007mj,Rosa:2011my,Cardoso:2011xi,Pani:2012bp,Destounis:2019hca,Vieira:2021xqw,Vieira:2021ozg,Vieira:2023ylz}, is the mechanism in hand that leads to the change of the fundamental mode when comparing a BH and an ECO of the same mass. The protrusion of the cases shown in Fig.~\ref{ECO_PS_nopert} occurs for $\epsilon\lesssim10^{-3}$, for the fundamental QNM, and $\epsilon\lesssim10^{-2}$ for the first overtone.

In order to further understand the protrusion's capability for an instability of the ECO under study, we calculate the perturbed QNMs of the system by adding random and deterministic perturbations $\delta L$ (with energy norm $\|\delta L\|\simeq \epsilon$) to the ECO potential, as was performed in~\cite{Jaramillo2020}. The fundamental mode always appears to be spectrally stable up to perturbations of energy norm $\epsilon\lesssim10^{-1}$ and higher, suggesting that a migration to the unstable side of the complex plane cannot be triggered. When the ECO is less compact and its surface reaches the photon sphere, then the perturbation response is significantly different. Pseudospectra are much harder to protrude into the unstable side of the complex plane for sufficiently small values of $\epsilon$. Nevertheless, the ECO QNMs are still spectrally unstable. From Fig.~\ref{ECOps}, it is apparent that the spectral stability of the fundamental ECO QNMs against random and deterministic perturbations is present only when their imaginary part is of the same, or smaller, order than the imaginary part of the fundamental Schwarzschild BH QNM.

Finally, it is important to notice the case where $\ell\gg 1$, specifically the case of scalar $\ell=10$ QNMs in Fig.~\ref{ECO_PS_nopert}, and how it connects to the fact that ECO instabilities are triggered in the parametric region of high angular modes. In the mentioned figure, we observe at least four QNMs (the fundamental and the first three overtones) that protrude to the unstable part of the complex plane. This suggests a stronger tendency to trigger pseudoresonances, which may make the ultracompact horizonless object unstable when nonlinearities are considered, as already shown to occur~\cite{Cardoso:2014sna,Keir:2014oka}.

This further justifies the strength of the pseudospectrum to potentially predict nonlinear transient effects even though the operator in hand is still linear. In particular, the transitive protrusion of pseudospectral contours through the real axis, along with the notion of pseudoresonances, can be combined with BH perturbation theory in GR at second order~\cite{Cunningham1980,Gleiser:1996yc,Miller:2016hjv} to give rise to a possible bootstrap instability mechanism, discussed in~\cite{Boyanov:2022ark} and exhaustively detailed in~\cite{Jaramillo:2022kuv}. Such a mechanism seems perfectly suited for ultracompact objects since pseudoresonances may lead to the growth of external time-dependent sources close to protrusions, which break down the perturbative expansion and leads to nonlinear dynamical instabilities. Therefore, in the gravitational cases studied in this section, the bootstrap instability mechanism is provided by the GR perturbative structure that sources second-order gravitational perturbations through first-order gravitational ones~\cite{Boyanov:2022ark,Jaramillo:2022kuv}, thus the characterization as a \emph{bootstrap mechanism}.

\section{Destabilizing the fundamental quasinormal mode}

In the previous sections, the pseudospectra of BHs and ECOs were studied, together with the migration of QNMs when the effective potential of the aforementioned compact objects was perturbed with random or deterministic (sinusoidal) perturbations. Random perturbations still do not possess any reasoning on why they may be important physically, but sinusoidal ones can be interpreted as quantum-gravity oriented, though this aspect is still up for discussion. Nevertheless, both perturbations introduced to the BH or ECO potentials led to the same conclusion: even though the pseudospectrum suggests that the fundamental mode can be destabilized by generic perturbations, actually perturbing only the effective potential would lead to the migration of overtones, while the fundamental mode was always remaining intact. Therefore, the fundamental QNM seems to admit a special status when it comes to perturbing the BH potential; a characteristic that is quite reassuring for BH spectroscopy.

The simplistic approaches that regarded BHs as isolated objects in vacuum or in binaries made them perfect laboratories for tests of GR, as long as their environment was ignored. Recent works~\cite{Cardoso:2021wlq, Cardoso:2022whc, Speri:2022upm, Speeney:2022ryg} have shown that not taking into account environmental effects, as well as other astrophysical uncertainties, may compromise precision tests of GR, especially when the shadow detections of dark compact objects in the galactic core of M87~\cite{EventHorizonTelescope:2019dse} and the Milky way~\cite{EventHorizonTelescope:2022wkp} solidified the existence of extremely rich environments around BHs, such as accretion disks~\cite{Lynden-Bell:1969gsv,Gurzadian:1979,Armitage:2004ga,Dittmann:2019sbm,Speri:2022upm}. Moreover, the suggestion that large length-scale, galactic dark matter halos act as ``galactic glue'' and contribute further non-luminous mass to the galaxy~\cite{Sofue:2008wt,Catena:2009mf,Weber_2010} made even more clear the need for taking into account both short and large-scale environmental effects when performing precision GW astrophysics. In this section, we will explore the possibility of not ignoring environmental effects, and in particular if they can lead to spectral instabilities in Schwarzschild BHs surrounded by matter shells.

\begin{figure}[b!]\centering
	\includegraphics[scale=0.45]{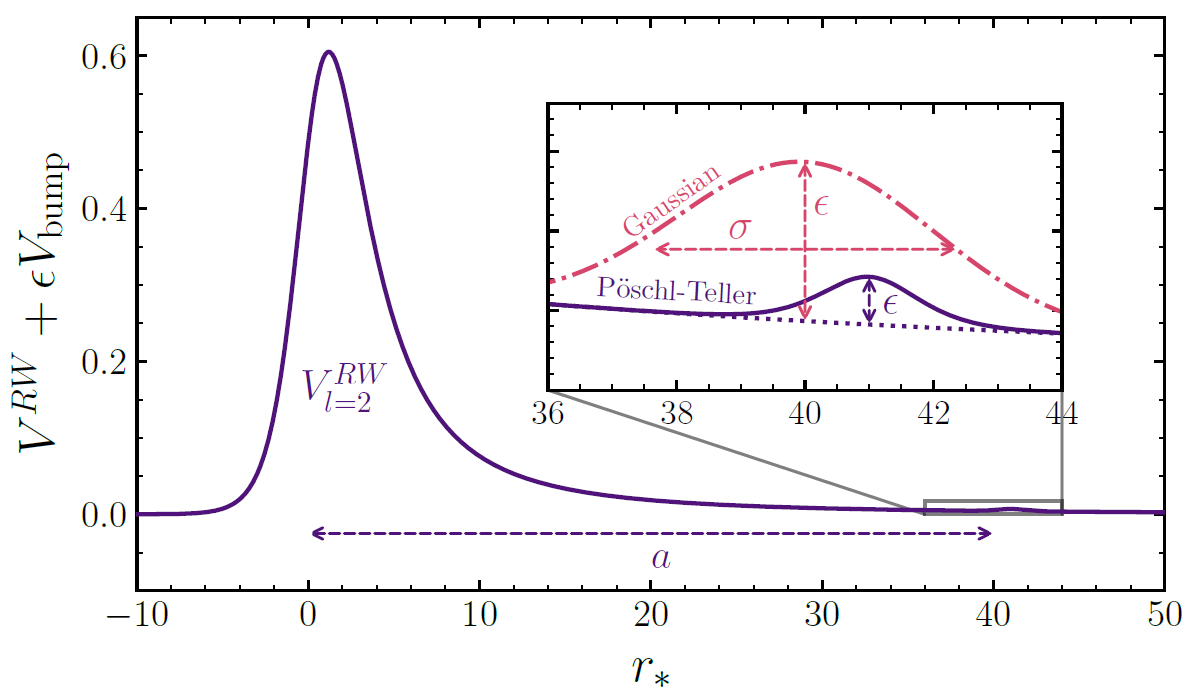}
	\caption{Schematic illustration of the $\ell=2$ axial gravitational (Regge-Wheeler) potential $V^{RW}$ perturbed by a P\"oschl-Teller bump (solid line in the inset) and by a Gaussian bump (dot-dashed line). The figure has been taken from~\cite{Cheung:2021bol}.}\label{potential_bump}
\end{figure}

To model the ringdown signal of a BH merger, by using QNMs, we usually take into account a part of the spectrum obtained in vacuum. In order to consider also surrounding matter, we will add a bump in the effective potential beyond the photon sphere. For simplicity, we consider the effective potential of a Schwarzschild BH and add a small bump to it of the form~\cite{Barausse:2014tra,Cheung:2021bol}
\begin{equation}\label{Vpert}
	V_\epsilon=V+\epsilon V_\text{bump},
\end{equation}
with $\epsilon\ll 1$ and $V_\text{bump}$ a generic matter shell that induces a bump located at $r_*=a$, such that the bump decays faster that $V$. By solving the frequency domain equation~\eqref{wave_equation_frequency} with QNM boundary conditions, and replacing its potential with that of Eq.~\eqref{Vpert}, we obtain the perturbed QNMs $\omega^{(\epsilon)}_n$. The bump drives the original, unperturbed, fundamental mode $\omega^{(0)}_0$ to migrate continuously in the complex plane along a curve that depends on the distance between the BH potential peak and the bump $a$. To avoid confusion, we will drop the subscript $0$ from the perturbed QNMs since we do not necessarily know if they are fundamental anymore after their migration. We further define the variation $\Delta\omega^{(\epsilon)}=|\omega^{(\epsilon)}-\omega^{(0)}|$. When the quantity $|\Delta\omega^{(\epsilon)}/\omega^{(0)}|\gg\epsilon$ then we either observe a destabilization of the fundamental mode due to disproportional migration in the complex plane or due to the phenomenon of \emph{mode overtaking}, where unperturbed overtones become fundamental modes as $a$ increases. In what follows, to avoid confusion we will denote the perturbed fundamental mode as $\varpi$, no matter if it is the original, the overtaken or the destabilized QNM.

We consider the potential $V$ to be that of $\ell=2$ axial gravitational perturbations on a Schwarzschild BH with mass $M=1/2$ and two types of bumps added in distance $a$, namely the P\"oschl-Teller potential
\begin{equation}
	V_\text{PT}(r_*-a)=\text{sech}^2(r_*-a),
\end{equation}
and a Gaussian peak 
\begin{equation}
	V_\text{G}(r_*-a)=\exp\left(-\frac{(r_*-a)^2}{2\sigma}\right), \label{eq:GaussianBump}
\end{equation}
where $\sigma$ is the Gaussian distribution's width. An illustration of the ``toy model'' potential is demonstrated in Fig.~\ref{potential_bump} \footnote{We stress that even though here we add the bump by hand in the BH potential, a matter shell can induce a bump with the same properties as those we observe here. See the Appendix A of Ref.~\cite{Cheung:2021bol} for a more detailed discussion.}.

\begin{figure*}\centering
	\includegraphics[scale=0.75]{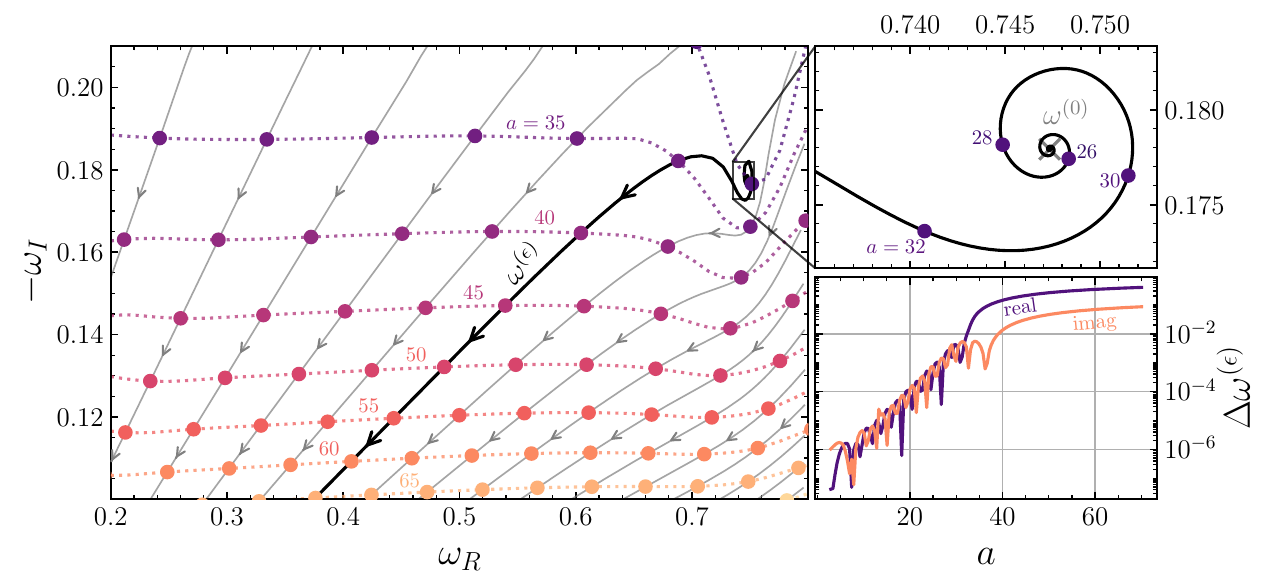}
	\caption{Migration of $\ell=2$ axial gravitational QNMs as a function of the bump position $a$ for a P\"oschl-Teller perturbation with $\epsilon = 10^{-6}$. Left: Individual modes migrate along the black lines. The mode $\omega^{(\epsilon)}$ (bold line) reduces to the Schwarzschild fundamental mode when $\epsilon\to 0$, and the arrows indicate the direction of migration as $a$ increases. Modes with the same value of $a$ have the same color, and are connected with dotted lines. Top right: Zoom-in around the unperturbed fundamental QNM $\omega^{(0)}$. Bottom right: Real and imaginary parts of the migration distance $\Delta\omega^{(\epsilon)}$ of the perturbed fundamental QNM. We use units such that $2M=1$. The figure has been taken from~\cite{Cheung:2021bol}.
	}\label{QNM_migration1}
\end{figure*}

The migration of QNMs due to a P\"oschl-Teller bump with $\epsilon=10^{-6}$ is shown in Fig.~\ref{QNM_migration1}. The left panel depicts the displacements of QNMs in the complex plane as we increase $a$. Specifically, the bold continuous curve traces the original fundamental mode's migration. In the top right panel, we zoom in on the trajectory in the close vicinity of $\omega^{(0)}$ for small to moderate $a$, while the bottom right panel shows the change in the variation of the real and imaginary parts of the perturbed QNM. The variation exceeds the perturbation scale by at least four order of magnitude, so we can conclude that even the fundamental mode of BHs can be destabilized, as conjectured by the pseudospectrum, by including appropriate and astrophysically-relevant perturbations to the effective potential.

\begin{figure*}\centering
	\includegraphics[scale=0.58]{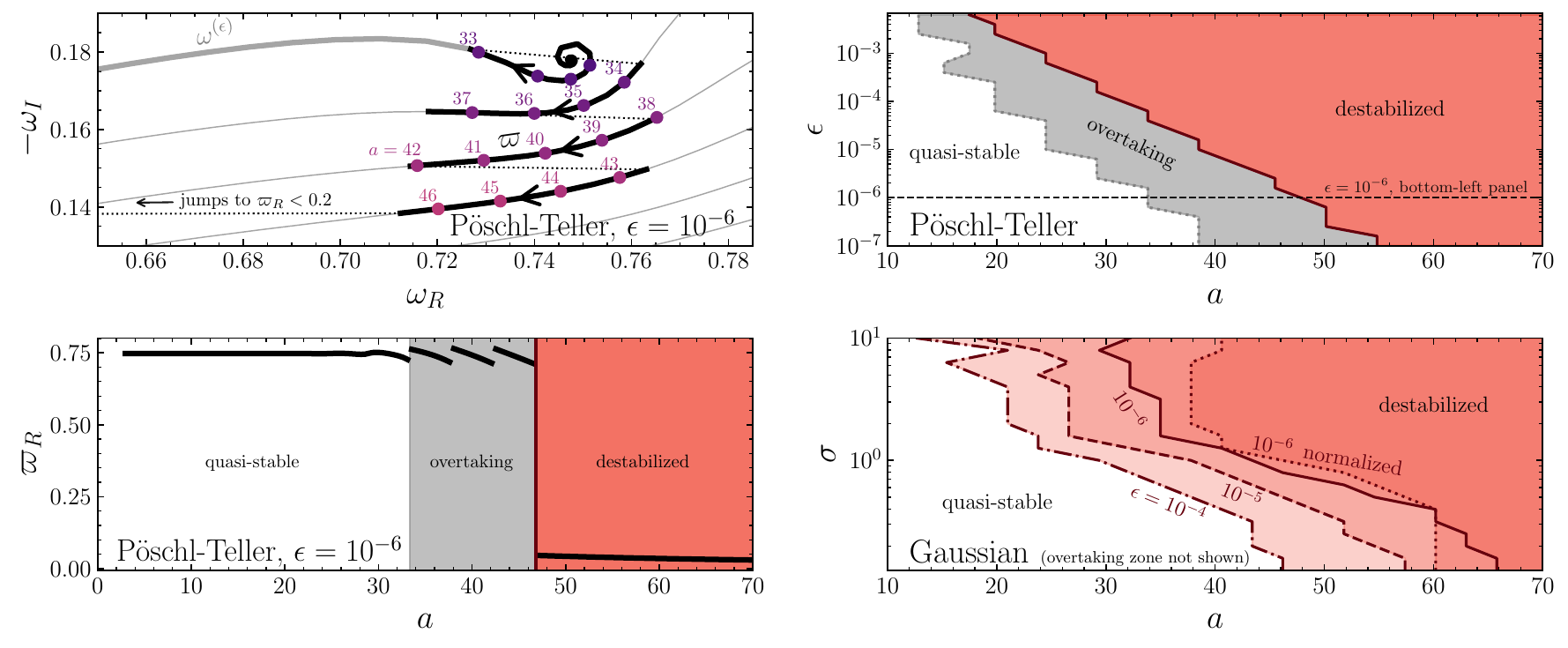}
	\caption{Top left: Migration of the fundamental mode $\omega^{(\epsilon)}$ in the complex plane when $a$ is increased, for a P\"oschl-Teller bump with $\epsilon = 10^{-6}$. Discontinuous jumps in $\omega$ (in the plot denoted as $\varpi$ which designates the new perturbed fundamental QNM) are shown with dotted lines. 
		Bottom left: Variation of the real part of $\varpi_R$  from the top left panel as a function of $a$. Top right: Phase diagram of $\epsilon$ vs $a$ for a P\"oschl-Teller bump. Bottom right: ``Phase diagram'' of $\sigma$ vs $a$ for a Gaussian bump with different values of $\epsilon$. The figure has been taken from~\cite{Cheung:2021bol}.
	}\label{phase_diagrams}
\end{figure*} 

Figure~\ref{phase_diagrams} presents the whole parametric analysis performed in~\cite{Cheung:2021bol}, where not only $a$ but also $\epsilon$ and $\sigma$ is varied, together with a different perspective on mode migration and overtaking by overtones. The top left panel details the discontinuous overtaking of the original fundamental mode. For small $a$, the fundamental mode remains the same even though it continuously migrates in the complex plane. Around $a\sim 30M$, an overtone $\omega^{(\epsilon)}$ takes over discontinuously and becomes the new fundamental mode $\omega^{(0)}$. After some further overtakings, the original fundamental mode has been completely destabilized and migrated up to order $\mathcal{O}(10^{-1})$. The bottom left panel demonstrates this behavior of migrating QNMs and how the overtaking occurs at the real part of the QNM frequency. The first regime, shown in white, consists of a quasi-stable region where the original fundamental mode has not been overtaken yet. In the grey region, three consecutive overtakings take place, where overtones become the new perturbed fundamental modes until the migration is so large that leads to a completely destabilized fundamental mode in the orange region. In the latter regime, the separation between the real parts of two consecutive modes is given to a very good approximation by the equality $\omega_{n+1,R}-\omega_{n,R}=\pi/a$. This is an expected characteristic of trapped modes between two potential barriers located at distance $a$ from each other and leads to repetitions of photon sphere excitations in the ringdown signal, i.e. echoes. Therefore, the overtaking and eventual destabilization of the fundamental mode is due to stable trapping and the appearance of longer-lived modes (with large decay timescales) that overtake the unperturbed QNMs, which decay faster than the trapped, long-lived ones.

We have repeated the analysis in a significant volume of the parameter space for P\"oschl-Teller and Gaussian bumps by varying $\epsilon, \sigma$ and $a$ in order to sketch phase diagrams that characterize the destabilization. In the right column of Fig.~\ref{phase_diagrams} we demonstrate that the larger $\epsilon$ is (for the P\"oschl-Teller case), the easiest is to form quasibound states between the two potentials since the destabilization takes place for shorter distances $a$. It is clear that no matter how small we choose the bump height to be, there is always a critical $a$ beyond which the destabilization of the fundamental QNM occurs. Similar results are shown for the Gaussian bump perturbation, where the wider the bump is, the easier is to destabilize the fundamental mode. From the above results, it is quite clear that the fundamental mode does not hold a distinguished status in the spectrum, as the analyses in~\cite{Jaramillo2020,Boyanov:2022ark} suggest, but can also migrate disproportionately as the pseudospectrum recommends. The question that now follows is if a destabilized fundamental mode, due to environmental effects, is present and dominant in the ringdown waveform of such a perturbed potential. We discuss this aspect in the following section.

\section{Destabilized fundamental mode detection in ringdown signals}

Until this point, we have been discussing the stability of the QNM spectrum from a \emph{formal} perspective, using the powerful tool of the pseudospectrum. We have concluded that, in general, the QNM spectra of BHs and ECOs in GR are spectrally unstable. However, what is the actual consequence of this for GW astronomy, and in particular for the BH spectroscopy program? In the presence of a perturbation in the effective BH potential, is the ringdown signal also destabilized in some sense? A key point in this question is that the QNM spectral instability is obtained in the frequency domain, while GW interferometers measure a strain in time. The two pictures should be equivalent and related by a simple Fourier transformation. However, this is only true if we have access to the full signal in time \emph{ad eternum}. In practice, this is impossible in an experiment due to systematics, such as instrumental noise. 

This situation is reminiscent to what happens in ECOs. Their QNM spectra are extremely different from a BH in GR, in particular, they have long-lived trapped modes. However, by causality, the time-domain response of a BH and an ECO is the same for the time necessary for a perturbation to travel to the surface of the ECO, which is close to the BH horizon limit, and then be reflected back~\cite{Cardoso:2016rao,Mark:2017dnq}. 

Here, we solve the equation governing axial gravitational perturbations in the time domain ($s=2$ in Eq.~\eqref{wave_equation_time}), using a two-step Lax-Wendroff method with hyperboloidal layers~\cite{Krivan:1997hc, Pazos_valos_2005, Zenginoglu:2007jw, Zenginoglu:2011zz, PanossoMacedo:2019npm} (we refer the reader to the references mentioned for more details on this numerical scheme). We focus on the case where the perturbation to the effective potential is described by a Gaussian bump, as in Eq.~\eqref{eq:GaussianBump}. However, our qualitative conclusions are independent of the actual shape of the perturbation. We considered other ones, such as a concave bump with negative $\epsilon$; and a bump decaying as $r^{-3}$ at large distances, which is exactly zero for small enough $r$, mimicking a perturbation due to a thin shell of matter. The only relevant aspect of the perturbation is that it introduces a peak in the effective potential, otherwise, the instability is not present as seen in the previous section. As initial data for the perturbation, we prescribe a localized Gaussian pulse
\begin{equation}
	\Psi \Big|_{t=0} = 0 \quad , \quad 
	\frac{\partial \Psi}{\partial t} \Big|_{t=0} = e^{-(r_* - 5)^2/2} \, , \label{eq:id_gaussian}
\end{equation}
but, again, our qualitative conclusions are independent of this particular choice.

Finally, in order to extract the ringdown spectra, we fit the time-domain waveform to a superposition of exponentially damped sinusoids
\begin{equation}
	\Psi(t) = \text{Re} \sum_{n=0}^{N-1} A_n e^{-i (\omega_n t - \phi_n)} = \sum_{n=0}^{N-1} A_n e^{\omega_{n I} t} \cos(\omega_{n R} t - \phi_n)\, ,
	\label{eq:template}
\end{equation}
where the index $n$ labels the different modes we find by fitting, and it does not necessarily coincide with the overtone number. Each mode is characterized by four parameters: an amplitude $A_n$, a phase $\phi_n$, and the real and imaginary parts of the QNM frequency $\omega_n = \omega_{nR} + i\omega_{nI}$.
We will find that several QNMs could have similar decay times and hence comparable amplitudes. When this happens, a good fit of the waveform requires a relatively large number of modes $N$. The largest number of modes we will look for is $N = 8$, corresponding to $8 \times 4 = 32$ fitting parameters.

\subsection{Ringdown waveform stability}\label{waveform_stability}

First, let us clarify what we mean by destabilization here. We will say that \emph{destabilization} occurs when a quantity that characterizes the BH's behavior under an external perturbation, such as its waveform amplitude, changes by an amount much larger than the magnitude of the perturbation (in this case introduced in the effective potential).

\begin{figure}[b!]
	\centering
	\includegraphics[width=16cm]{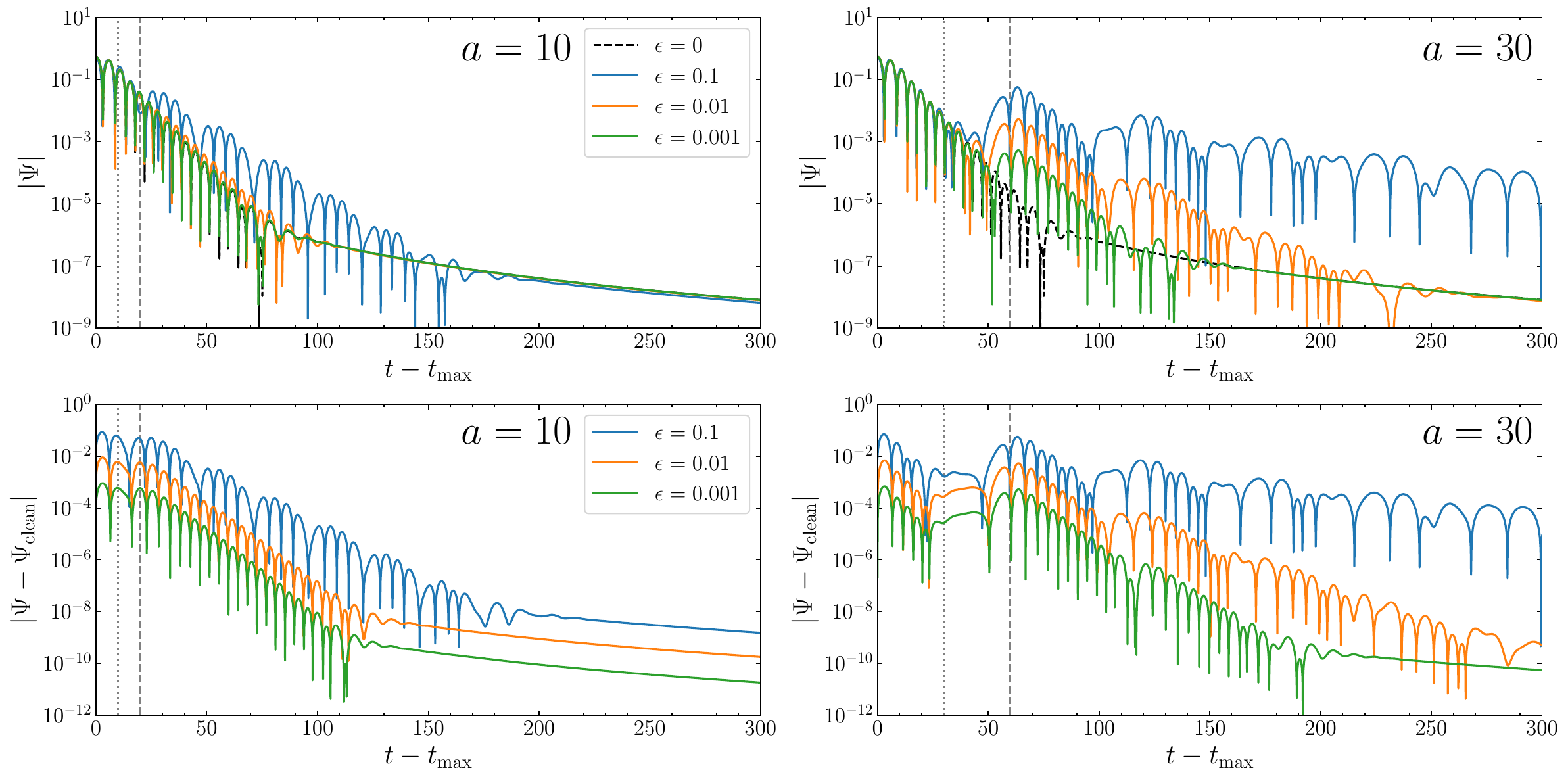} 
	\caption{Top panel: absolute value of the waveform arising from the scattering of the Gaussian pulse of Eq.~\eqref{eq:id_gaussian} for bumps with different amplitudes $\epsilon$, located at two selected distances $a$ from the main peak. The bump's width $\sigma$ in Eq.~\eqref{eq:GaussianBump} is fixed at $\sigma = 0.5$. Echoes are apparent when the bumps has a large separation from the photon sphere ($a = 30$). The dotted and dashed vertical gray lines correspond to $t - t_\text{max} = a$ and $2a$, and they illustrate how the delay between echoes is related to the size of the ``cavity'' between the two maxima in the perturbed potential. Bottom panels: Absolute value of the difference between the ringdown waveforms shown in the top panel and the unperturbed clean waveform without a bump ($\epsilon = 0$). The figure has been adapted from \cite{Berti:2022xfj}.}
	\label{fig:WaveformsEps}
\end{figure}

Figure~\ref{fig:WaveformsEps} shows the resulting ringdown waveform from the scattering of Gaussian pulses for different bumps. We define $t_{\rm max}$ as the instant at which $\left| \Psi \right|$ is maximum. In Schwarzschild ($\epsilon = 0$, black curve), one obtains the typical exponentially damped quasinormal ringing followed by the power-law tail due to backscaterring of radiation on the background curvature~\cite{Leaver:1986gd}. The larger the $\epsilon$, the larger the difference in the waveform with respect to vacuum. For earlier times, when $t-t_{\rm max} \lesssim a$, the absolute difference on the waveform with respect to the unperturbed potential case scales linearly with $\epsilon$. The \emph{prompt} ringdown signal close to the waveform peak is thus not destabilized in the sense defined above. This is the portion of the ringdown with the most interest to GW astronomy, meaning such differences are not expected to be observable with the singal-to-noise ratios (SNRs) achievable by current interferometers~\cite{Barausse:2014tra,Cardoso:2019rvt}.

\begin{figure}[t]
	\centering
	\includegraphics[width=0.6\linewidth]{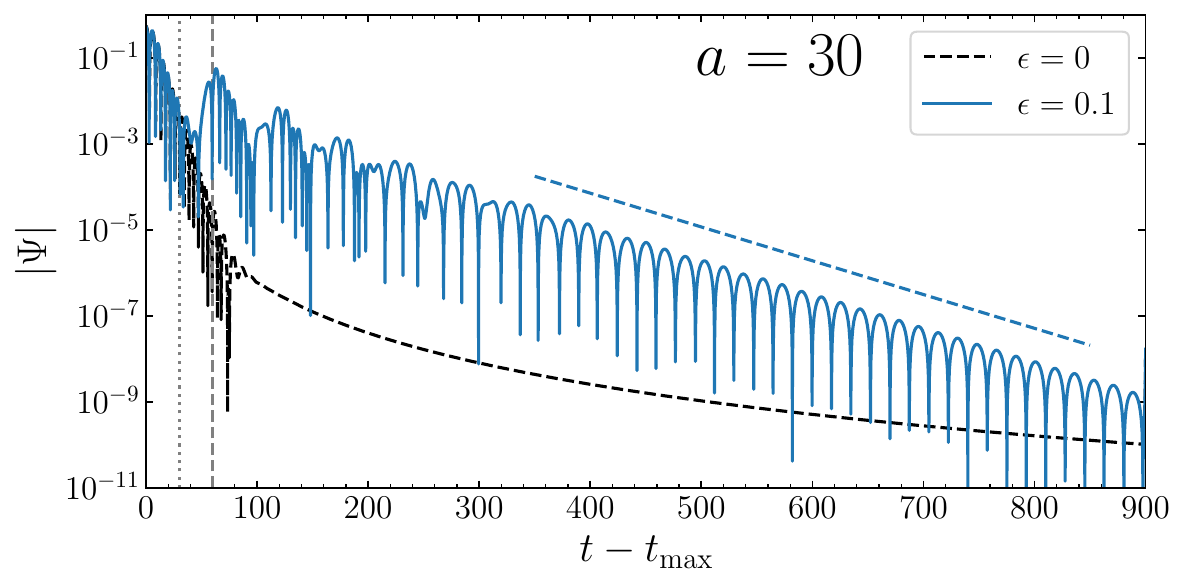} 
	\caption{Ringdown waveform for $a=30$ and $\epsilon = 0.1$ over long time windows, where the new perturbed fundamental QNM dominates the late-time behavior of the signal. The dashed blue line represents its expected decay, which corresponds to the bottom blue cross with smallest $\left| \omega_{I}\right|$ in the top panel of Fig.~\ref{fig:FitCompFrequenciesa10}. The figure was taken from \cite{Berti:2022xfj}.}
	\label{fig:LongWaveform}
\end{figure}

After this initial regime, for $t-t_{\rm max} \gtrsim 2a$, we observe the repetition of the original prompt ringdown. These are echoes of waves trapped between the two peaks of the effective potential that ``leak'' out to infinity. At late times, the signal's frequency content can be drastically different. Here, the waveform is well described by a superposition of the long-lived QNMs of the perturbed spectra, as illustrated in Figure~\ref{fig:LongWaveform}. For smaller $\epsilon$, the power-law tail can hide the difference in the late-time behavior with respect to the Schwarzschild waveform. 

Typically,  larger bumps lead to QNM spectra with longer damping times, which thus decay slower and survive longer before being dominated by the power-law tail. For large $a$ there is a hierarchy in the timescales between the prompt ringdown, which is excited at the photon sphere, and the travel time of waves characterizing the \emph{cavity} between the two peaks of the effective potential, located at the photon sphere and at the bump. Thus, a pulse bounces back and forth within the cavity and gradually loses its high-frequency component, which tunnels out more easily. This produces a sequence of echoes repeating at a characteristic frequency defined by the cavity size and damped on a timescale defined by the transmission coefficient of the small peak, as shown in Figure~\ref{fig:WaveformsEps} (see also Refs.~\cite{Cardoso:2016rao,Cardoso:2016oxy,Cardoso:2017cqb,Cardoso:2019rvt} for similar behavior when the bump is arbitrarily close to the horizon). These two scales determine the QNM spectrum of the bumpy potential, which is unstable and therefore can be non-perturbatively different from the $\epsilon=0$ case.

The prompt ringdown is excited mainly at the peak of the effective potential, which is close to the photon sphere. If we place the bump closer to it, it changes the shape of the peak and, consequently the frequency content of the prompt ringdown also. If the bump is farther, the spectrum also changes because the QNMs are sensitive to the entire potential. However, the wave train excited at the peak should be similar to the one in the case of the unperturbed potential. Part of it will be reflected as it meets the bump, and the other will tunnel out to far-away distances. The reflection coefficient is of order $\mathcal{O}(\epsilon)$ for any frequency, and consequently, the change in the prompt ringdown also scales with this factor, designating ringdown stability at early times.    

In the prompt ringdown, the relative difference in the waveform $\left|\Psi - \Psi_\text{clean} \right| / \left| \Psi \right|$ scales linearly with $\epsilon$, while in the echo-dominated regime the difference is larger. The reason for this is the following: the original ringdown signal decays as $e^{-\omega_I t}$. Each reflection of the waves in the cavity reduces their order of magnitude by $\sim\epsilon$. On the other hand, each back-and-forth bounces inside the cavity occurs on timescales of $t_\text{bounce}\sim 2a$. This means that the amplitude of the \emph{n}-th echo will be larger than the ringdown by a factor of $\left(\epsilon / e^{-\omega_I t_\text{bounce}}\right)^n$, or $\epsilon^n e^{n2a \left|\omega_I\right|}$. When the power-law tail starts dominating, the modification returns to order $\epsilon$ because the tail can be formally seen as a ``direct'' zero-frequency signal.        

\subsection{Implications on black-hole spectroscopy}

Having understood the stability problem regarding the ringdown, we focus on the core procedure of BH spectroscopy, which is to compare QNM spectra of compact objects with the frequencies extracted by fitting the ringdown with damped sinusoids. However, there are two different questions:

\begin{enumerate}
	\item How the spectral instability affects the prompt ringdown, which is the louder portion of the signal and consequently of more interest for GW astronomy.
	
	\item If the full waveform, in particular the late-time content, is well described by the destabilized QNM spectrum, including the long-lived trapped modes.
\end{enumerate}

\begin{figure}[t]
	\centering
	\begin{tabular}{c}
		\includegraphics[width=0.6\linewidth]{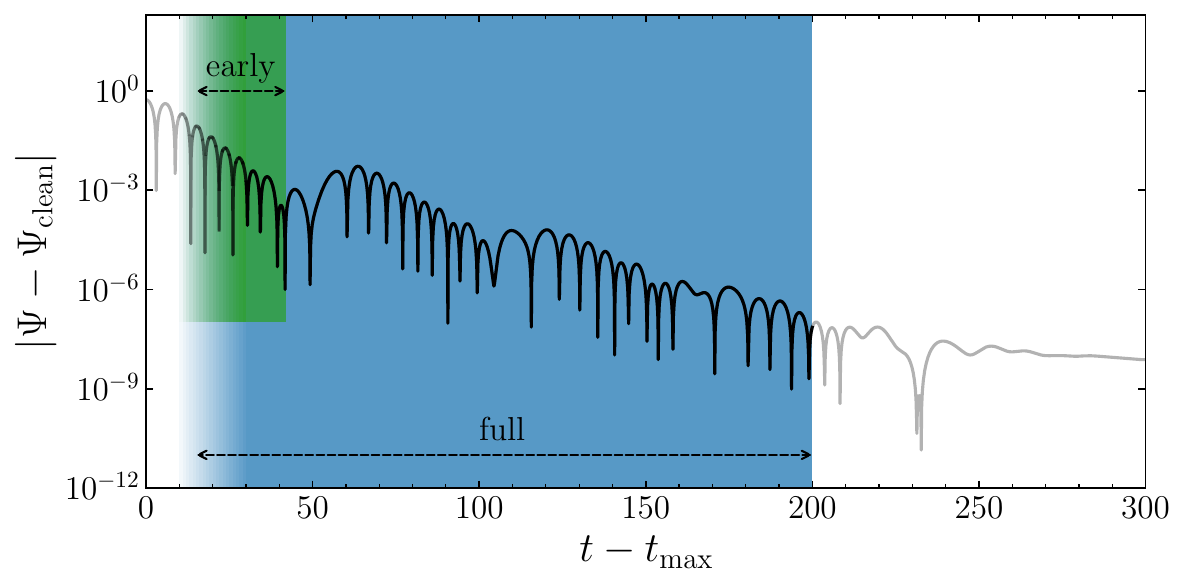}  
	\end{tabular}
	\caption{The portion of the waveform used for the damped-sinusoid fitting in the two different regimes of interest. For completeness, this waveform corresponds to the case $a=30, \, \epsilon =0.01$, but the same procedure applies to other examples. The figure has been taken from~\cite{Berti:2022xfj}.}
	\label{fig:FitRange}
\end{figure}

To answer these, we use two different portions of the ringdown signal when fitting the time-domain waveforms. These are highlighted in Figure~\ref{fig:FitRange} . In both cases, we discard times $t-t_\text{max}\lesssim 5$, where there is contamination from the direct signal coming from the initial data~\cite{Baibhav:2023clw}. When studying the full signal (shaded in blue), we discard the portion of the waveform dominated by the power-law tail. For the prompt ringdown analysis (shaded in green), we only consider the portion of the signal before the appearance of the first echo ($t-t_\text{max} \lesssim a$). In every fit, the starting time is varied to ensure convergence of the extracted frequencies.

\subsubsection{The full time-domain signal}

\begin{figure}[htb]
	\centering
	\begin{tabular}{c c}
		\includegraphics[width=0.4\linewidth]{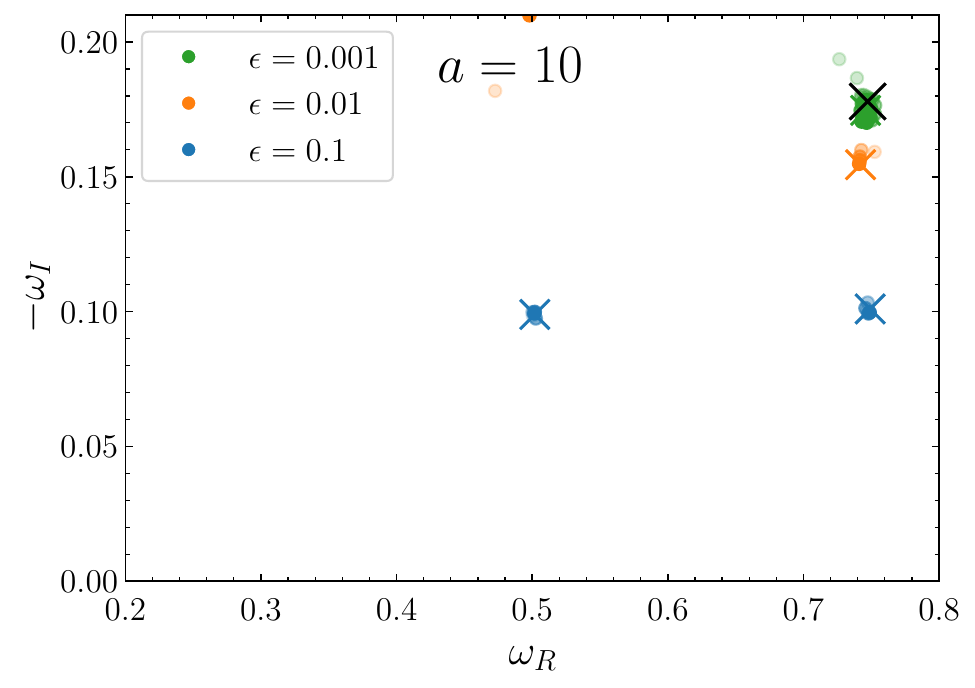} 
		\includegraphics[width=0.45\linewidth]{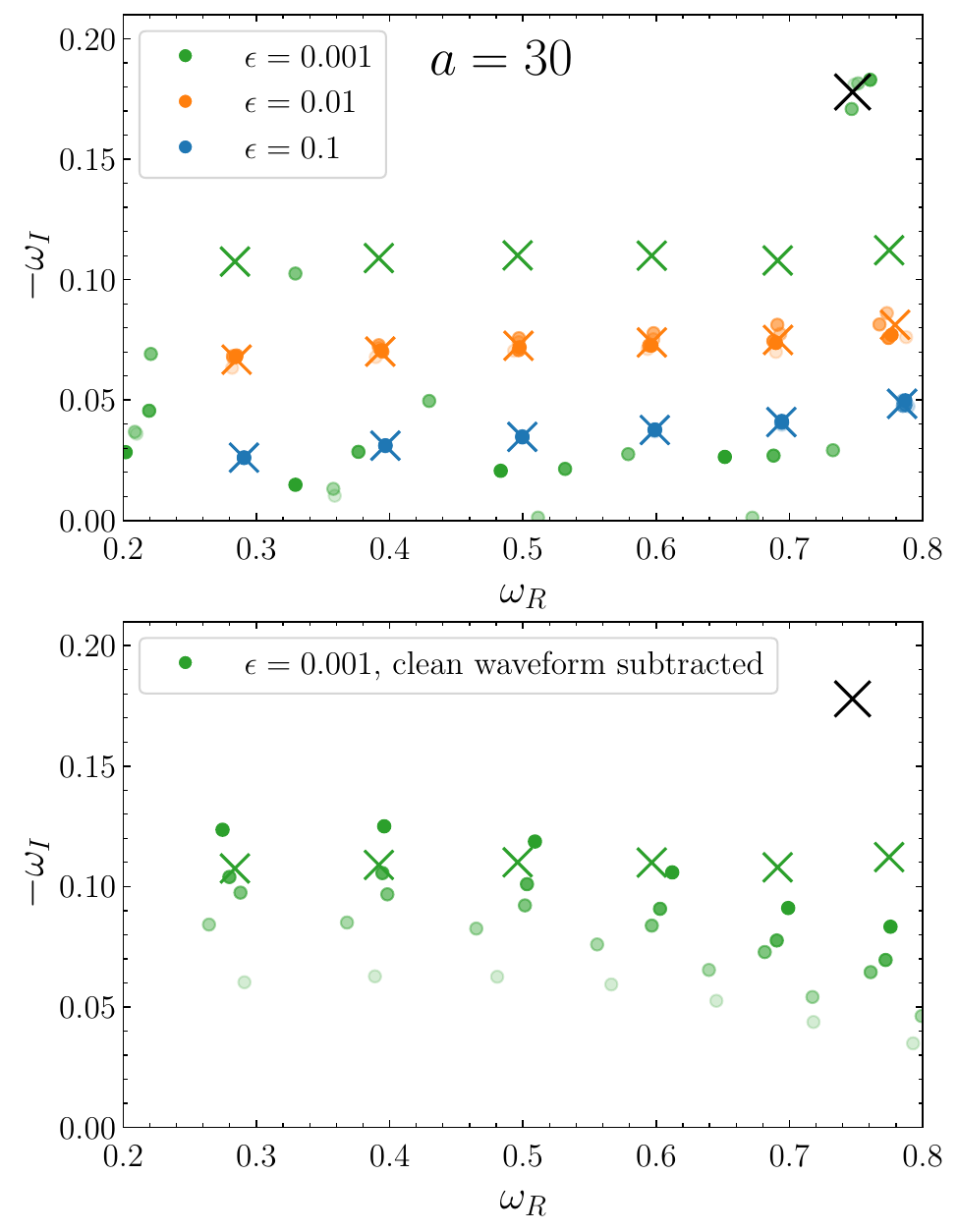}
	\end{tabular}
	\caption{Left panel: Comparison between the QNM frequencies computed with the shooting method in the frequency domain (crosses) and those extracted by fitting the full time domain waveform (dots). The black cross corresponds to the unperturbed clean fundamental QNM of Schwarzschild, where $M\omega=0.374 - 0.089 i$. The starting times used are $t-t_\text{max} = 10, \, 15, \, 20, \, 25, \, 30M$. Darker shaded dots correspond to later initial fitting times. Right top panel: Same as in the left panel but with $a=30$. For $\epsilon = 0.001$, the full time-domain fits can only confidently detect a mode close to the fundamental mode of Schwarzschild, because the power-law tail dominates the signal before it transitions to the new set of perturbed QNMs. Right bottom panel: Same analysis as in the right top panel for $\epsilon = 0.001$, but subtracting the unperturbed clean waveform from the full signal before performing the fit. The frequencies extracted do not converge exactly to the QNMs, but the structure is more similar, in particular for the real part. The figure is adapted from~\cite{Berti:2022xfj}.}
	\label{fig:FitCompFrequenciesa10}
\end{figure}

We begin by addressing the problem of the full signal. Figure~\ref{fig:FitCompFrequenciesa10} demonstrates results for bumps located at $a=10$ and $a=30$. The fitted (extracted) frequencies from the time-domain signals are shown with dots with different shades, where darker dots correspond to later starting times. The crosses, instead, are the QNM frequencies computed in the frequency domain, with the black cross denoting the fundamental axial QNM of Schwarzschild ($\epsilon = 0$) and the rest of the crosses denote the QNMs for specified bump sizes $\epsilon$. We observe that the fitted frequencies for the bump closer to the potential peak ($a=10$) are in very good agreement with the ones predicted by the frequency-domain computations for all values of $\epsilon$ presented, when the fitting time starting point is larger than the prompt response time, where we avoid initial data contamination. The minor discrepancies can be attributed to numerical error and contamination by the initial data and the power-law tail. For the further bump ($a=30$), the destabilization of the spectrum is more noticeable. In this case, the new perturbed QNMs decay slower. After one echo, the waveform transitions to a combination of new QNMs until their amplitude becomes so small that they are masked by the late-time power-law tail. For $\epsilon = 0.1$ the ringdown can even transition to a clean exponential decay controlled by the new perturbed fundamental mode, as illustrated in Fig.~\ref{fig:LongWaveform}. This allowed us to extract multiple slowly decaying modes from the fit when $\epsilon \gtrsim 0.01$.

The fitting procedure becomes more complicated when $\epsilon \lesssim 0.001$. For these parameters, we can only confidently claim the presence of a mode close to the clean fundamental mode of the unperturbed potential. This is a consequence of the very short decay time of the QNMs for small perturbations and therefore the waveform does not transition to the new set of QNM spectra before the power-law tail dominates the signal. In an attempt to try removing the contribution of the tail, in the bottom panel of Fig.~\ref{fig:FitCompFrequenciesa10}, we subtract the clean ($\epsilon=0$) waveform from the signal and repeat the fit using the green curve in the bottom-right panel of Fig.~\ref{fig:WaveformsEps}. There are additional QNM oscillations that were previously hidden by the power-law tail. The fitted modes do not converge as when $\epsilon \geq 0.01$, but their structure is in better agreement with the predicted QNM spectrum. The upshot is that if the perturbation is strong enough (typically for larger $\epsilon$/smaller $a$), we can safely recover the unstable spectrum by fitting the full time-domain signal to a collection of modes, even though as we demonstrated in Section \ref{waveform_stability}, the waveform itself is stable under perturbations.


\subsubsection{The prompt ringdown}

\begin{figure}[htb]
	\centering
	\begin{tabular}{c}
		\includegraphics[width=0.85\linewidth]{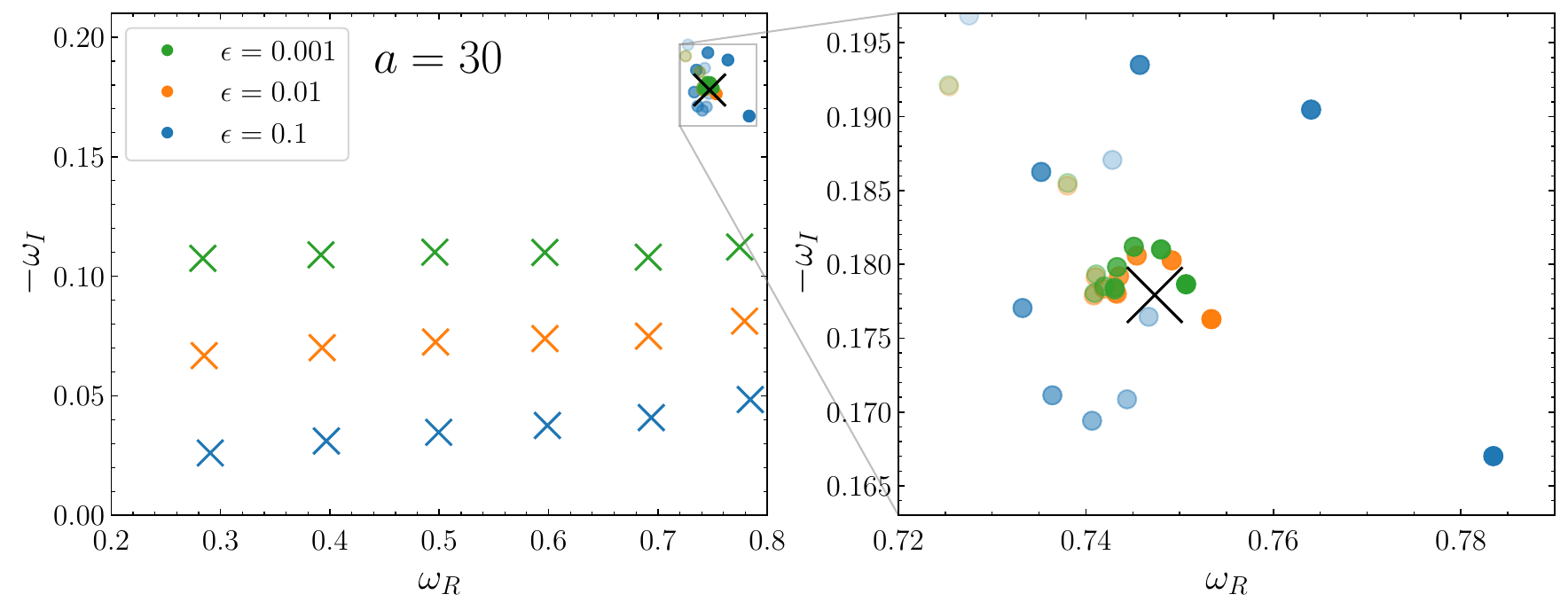}  
	\end{tabular}
	\caption{Same as Fig.~\ref{fig:FitCompFrequenciesa10}, but we only fit the first train of the initial ringdown without echoes. The starting times are $t - t_\text{max} = 10, \,11,\, 12,\, \dots,\, 20M$. All the dots obtained using the fitting method now cluster around the clean fundamental mode of the unperturbed potential. The zoom-in in the right panel shows that a perturbation of order $\epsilon$ can induce systematic errors (approximately of order $\epsilon$) in the measurement of the fundamental mode's frequency and damping time. The figure has been taken from~\cite{Berti:2022xfj}.}
	\label{fig:FitCompPrompt}
\end{figure}

In a real GW detector, we do not have access to the full signal due to the presence of noise. In addition, for astrophysical systems, the perturbation bump should be small. Matter typically introduces corrections in the effective potential of amplitude $\epsilon V_\text{bump} \sim \rho$, where $\rho$ is the matter density~\cite{Kokkotas:1999bd,Cardoso:2021wlq,Cheung:2021bol}. This corresponds to
\begin{equation}
	\rho M^2 = 1.6 \times 10^{-18} \frac{\rho}{\rho_{\rm water}}\frac{M^2}{M_\odot^2 } \, , 
\end{equation}
where $\rho_\text{water}$ is the density of water, $M$ is matter's mass and $M_\odot$ the solar mass. Therefore, we expect $\epsilon V_\text{bump}$ to be small in most astrophysical scenarios. Thus, with both current and future SNRs, we should only have access to the prompt ringdown signal. 

Because of this, in Fig.~\ref{fig:FitCompPrompt} we repeat the analysis of the previous section but restrict it to the \textit{prompt} portion of the signal before the appearance of the first echo, for times $t-t_{\rm max} \lesssim a$, shaded in green in Fig.~\ref{fig:FitRange}. Remarkably, the prompt ringdown is well described by a \emph{single} mode with a frequency close to the unperturbed fundamental QNM of Schwarzschild, in contrast with the vast spectrum recovered when fitting the full signal. Thus, we are led to conclude that the observed frequency associated with the prompt ringdown is spectrally stable, in a similar spirit to what happened with the waveform. The combination of our results for the full and prompt signal implies that the BH spectroscopy program is not jeopardized by the instability of the QNM spectra of very compact objects in GR. In order to recover the \emph{correct} destabilized QNM spectrum obtained in the frequency domain, one needs to consider the late-time portion of the signal, which is not expected to be observable with both current and future GW detectors. Even this might not be enough, as the power-law tail can dominate the signal's amplitude, which would hide the presence of new perturbed modes. 

Our conclusions highlight the necessity of always complementing computations of the QNM spectra in modified gravity with time-domain studies in order to verify that changes are present in the prompt ringdown. We have to note, however, that the analysis presented here was restricted to the spectral instability of the fundamental QNM. As discussed in the previous sections, overtone destabilization is more severe for short-wavelength deterministic perturbations that we have focused on in this discussion, and these overtones generally can affect the signal at early times and the predictions of BH spectroscopy~\cite{Jaramillo:2021tmt}. Yet, as we also mentioned, the extraction of overtones from waveforms is highly nontrivial, even at the linear level, and is a subject of current research and debate~\cite{Isi:2021iql, Cotesta:2022pci, Baibhav:2023clw}. Any future investigation of the impact of overtone spectral instabilities in GW signals must be complemented by a better understanding on how to fit them accurately.

\section{Conclusions and outlook}

BH spectroscopy is the pinnacle of GW ringdown physics. By taking into account recent improvements concerning nonlinearities at the early ringdown stage, and by eventually finding a common ground regarding the status of overtone detection, we will soon be able to perform unprecedented tests regarding the validity of GR, the existence and properties of environments and galactic halos, as well as the quantum nature of spacetime. Nevertheless, there are still many aspects of the BH spectroscopy program that need improvement in order to form a trustworthy tool, when future ground- and space-based interferometers begin their observation runs, in order to analyze GWs from a huge variety of radiation sources in the cosmos. 

In this chapter, we have portrayed yet another interesting aspect of QNMs, and how it may affect BH spectroscopy, namely the fact that they suffer from spectral instabilities. This practically means that considering BH and ECO solutions only in vacuum, which is currently being done with numerical relativity simulations to evolve compact object binaries, may spoil precision GW astronomy. In order to have more accurate models of waveforms from coalescing binaries to perform data analysis and parameter estimation, environmental effects must be taken into account. Considering non-vacuum binaries comes with a drawback; the environment may taint QNMs and ringdown signals in accord with its nature and how strongly it affects the vacuum results.

We have described the current literature of spectral instabilities that occur in BH and ECO QNMs at the frequency domain with the mathematical notion of the pseudospectrum. We demonstrated that static, spherically-symmetric, and asymptotically-flat BHs exhibit an overtone instability, even including the fundamental mode when matter shell environments are taken into account in the effective potential of linear field perturbations. At the level of the spectrum itself, these results show a fragility in modal analysis, and if we want to further comprehend the underlying structure of QNMs we need non-modal tools, like the pseudospectrum, to gain more intuition into the limitations of linear predictions and transient phenomena that are obscured with typical eigenvalue investigations. Specifically for horizonless ECOs, we showed that the pseudospectrum may point towards a linear mechanism that could possibly lead to nonlinear instabilities when the ECO is ultracompact. This nonlinear instability mechanism is extremely interesting when one considers that the QNM operator is linear.

As mentioned above, if only the overtones suffer from spectral instabilities, then BH spectroscopy is safe at the moment since the fundamental mode is stable and plays a crucial role in analyzing ringdown signals. Unfortunately, it has been found that at the frequency domain, even the fundamental mode can be significantly destabilized with the addition of a tiny matter shell at some distance away from the photon sphere. This result may prove highly problematic for BH spectroscopy if such spectral instability is seeded into the ringdown waveform. Luckily, a time-domain analysis of linear gravitational perturbations on a Schwarzschild BH with a matter shell bump proves that the fundamental mode remains unsullied. The prompt ringdown waveform and the frequency fitted from this early part of the signal has an error proportional to the perturbation introduced in the potential. Only at very late times the spectrally-unstable fundamental mode appears, though our current detectors do not possess the SNR needed to probe such a long-lived mode. 

There is still plenty of work to be done for a spectroscopy program to be unequivocally successfully, especially when LISA begins its observation run of novel targets that can last for years and give much cleaner ringdown waveforms. Specifically, when the overtone dispute is solved, one must also perform analyses that include not only the stability of the fundamental mode in ringdowns, but also the overtones which play a crucial role in tests of GR and the no-hair theorems. Ultimately, the pseudospectrum should be vigorously added to the spectroscopy program while the holy grail of spectral (in)stability, namely the pseudospectrum of Kerr BHs, has to be examined in order to further understand if QNM instabilities are attributes of spherical symmetry or are omnipresent to all compact objects in the Universe without any symmetry assumptions.

\section*{Acknowledgments}	

The authors warmly thank the organization committee of the 11th Aegean Summer School that took place in Syros, Greece. The authors also want to warmly thank all the collaborators and colleagues that, even though not present at the Summer School, contributed significantly on the works discussed in this chapter, namely Emanuele Berti, Valentin Boyanov, Vitor Cardoso, Mark H. Y. Cheung, Edgar Gasperin, Jose-Luis Jaramillo, Rodrigo Panosso Macedo and Lamis Al Sheikh. 
K.D. acknowledges financial support provided by the DAAD program for the ``Promotion of the exchange and scientific cooperation between Greece and Germany IKYDAAD 2022'' (57628320). K.D. also acknowledges financial support under the European Union's H2020 ERC, Starting Grant agreement no. DarkGRA-757480 and the MIUR PRIN and FARE programmes (GW-NEXT, CUP: B84I20000100001).
F.D. acknowledges financial support by FCT under project No. 2022.01324.PTDC, and financial support provided by FCT/Portugal through the grant No. SFRH/BD/143657/2019.
%
\bibliography{Pseudospectrum}
\end{document}